\documentclass[10pt]{article}

\usepackage{amsmath}
\usepackage{amssymb}

\usepackage{graphicx}

\usepackage{cite}

\usepackage{color} 

\usepackage[labelfont=bf,labelsep=period,justification=raggedright]{caption}

\makeatletter
\renewcommand{\@biblabel}[1]{\quad#1.}
\makeatother

\date{}

\begin{document}

% Title must be 150 characters or less
\begin{flushleft}
{\Large
\textbf{Multidimensional epistasis and the transitory advantage of sex}
}
% Insert Author names, affiliations and corresponding author email.
\\
Stefan Nowak, 
Johannes Neidhart, 
Ivan G. Szendro,
Joachim Krug$^{\ast}$
\\
% \bf{1} 
Institut f{\"u}r Theoretische Physik, Universit{\"a}t zu K{\"o}ln, Cologne, Germany
\\
$\ast$ E-mail: krug@thp.uni-koeln.de
\end{flushleft}

% Please keep the abstract between 250 and 300 words
\section*{Abstract}
Identifying and quantifying the benefits of sex and recombination is a
long standing problem in evolutionary theory. In particular, contradictory claims have been made
    about the existence of a benefit of recombination on high
    dimensional fitness landscapes in the presence of sign epistasis.
 Here we present a comparative numerical study of sexual and asexual
 evolutionary dynamics of haploids on
 tunably rugged model landscapes under strong selection, paying special attention to the temporal development
 of the evolutionary advantage of recombination and the link between
 population diversity and the rate of adaptation. We show that the
 adaptive advantage of recombination on static rugged landscapes is
 strictly transitory. 
At early times, an advantage of recombination arises through the possibility to combine individually occurring beneficial mutations, but this effect is reversed at
longer times by the much more efficient trapping of recombining
populations at local fitness peaks.  
These findings are explained by means of well established results for
a setup with only two loci. In accordance with the Red Queen hypothesis the transitory advantage can be 
prolonged indefinitely in fluctuating environments, and it is maximal when the environment fluctuates on
the same time scale on which trapping at local optima typically occurs.

\section*{Author Summary}
Sexual reproduction is costly and often dangerous for the mating
individuals, but nevertheless it is wide spread in all higher organisms.
To resolve this paradox, evolutionary theory searches for fundamental
benefits in the recombination of genes that balance the cost of sex.
Although many possible benefits have been proposed, they often rely on the
assumption that each gene contributes independently to the fitness of an
individual.  Empirical studies indicate, however, that interactions
between genes are prevalent in nature and strongly constrain the process
of adaptation. Using a computational model, we show that these constraints
lead to an advantage of recombination that is effective only in the early
stages of adaptation to a new environment and turns into a disadvantage in
the long run. A sustained benefit of recombination can arise if the
environment is changing permanently, and we identify the conditions that
maximize this benefit.

\section*{Introduction}
Sexual reproduction is a phenomenon that has existed for more than
one billion years \cite{Butterfield2000}. It is widespread in nature, which means that
it must be advantageous compared to asexual reproduction, at least under certain conditions.
Nevertheless, it is not well understood why this should be the
case, since  sex has several major disadvantages, e.g., the \textit{two-fold cost of sex} \cite{maynardsmith1978,michod1987}, which
arises from the fact that two individuals are needed for reproduction without doubling the offspring in comparison to
asexuals. But even if one considers mere genetic recombination,
ignoring all the additional costs related to distinct sexes and
mating, the advantage of genetic reshuffling is far from obvious
\cite{Feldman1997,Otto2002,deVisser2007,Otto2009}. In particular, useful structures in the genome may be disrupted by
recombination giving rise to a \textit{recombination load} \cite{charlesworth1975}.

In order to explain the prevalence of sexual reproduction, or at least recombination, many theories
have been proposed. For example, \textit{Muller's ratchet} \cite{muller1964,felsenstein74} and the related \textit{deterministic mutation
hypothesis} \cite{kondrashov1988} explain the existence of sexual recombination through the difficulty of asexual reproduction to purge deleterious mutations. 
Other theories, like e.g., the \textit{Red Queen hypothesis}
\cite{valen1973}, claim that the advantage of recombination may not lie in its ability to facilitate higher fitness values in the long run.
Rather, it speeds up adaptation in order to be able to survive the
competition with other co-evolving organisms and changing environments. Again, various mechanisms are conceivable that may cause such a 
speedup of adaptation. Fisher \cite{fisher1930} and Muller \cite{muller1932}
first described what is now known as the \textit{Fisher-Muller effect} (which is
closely related to the \textit{Hill-Robertson effect} \cite{hillrobertson66,felsenstein74}): if in a population two beneficial mutations occur,
recombination of the two genomes can lead to
offspring harboring both mutations, while in asexual
populations the double mutant can be created only if 
one of the mutations reoccurs by chance on the background of the other
mutant.  
The \textit{Weismann effect}, on the other hand,
ascribes the advantage of sex to the creation of variation amongst siblings \cite{weismann1889}. The advantage of the Weismann effect is
implied in \textit{Fisher's fundamental theorem}, which states that the mean fitness gain of a population under natural
selection is proportional to its additive genetic variance \cite{fisher1930}. 

To be able to assess the effectiveness of the proposed mechanisms, it is necessary to make some simplifying assumptions.
In order to describe evolutionary processes and adaptation, Wright
introduced the notion of a \textit{fitness landscape} \cite{Wright1932}.
Fitness is a real number that is a measure
for the reproductive success of an organism, and the fitness landscape
is a mapping that assigns fitness values to genotypes \cite{deVisser2014}. Throughout the paper the term fitness landscape
  refers to this \textit{genotype landscape}, to be distinguished from
  the mean fitness landscape, also known as the adaptive landscape,
  which describes the mean fitness of a population as a function of
  its genotype or allele frequencies \cite{Provine1986,Buerger2000}.
Each site of the landscape corresponds to a
possible genotype and the individuals are distributed among these sites.
While mutation and recombination enable the exploration of formerly
unpopulated sites, selection leads to the concentration 
on genotypes with particularly high fitness. Thus, the population as a 
whole performs an uphill walk through the landscape and hence, on average,
an increase of fitness in time can be expected. 

In recent works, the  importance of \textit{epistasis} for the
evolutionary dynamics has been highlighted:
If the change of fitness associated with a mutation depends not only on the mutation itself, but also on the configuration of the
rest of the genome, the fitness landscape is called \textit{epistatic}
\cite{deVisser2011}.  
If the fitness effect of a mutation is altered by
the state of the other loci only in magnitude but not in sign, the effect is called \textit{magnitude epistasis}. If a mutation can be
beneficial or deleterious depending on the rest of the genome, the
effect is called \textit{sign epistasis} \cite{Weinreich2005}. 
Epistasis affects the mutational accessibility \cite{Poelwijk2007,franke2011,Franke2012,Nowi2013,Schmiegelt2014} of
fitness landscapes as well as the number of fitness optima, i.e.,
genotypes that have higher fitness than all of their single-mutant
neighbors \cite{Whitlock1995,Poelwijk2010,Crona2013}, and the length of adaptive
walks leading to such states 
\cite{gillespie1983,flyvbjerg1992,orr1,orr2003,joyceorr,krugneid,jain2011}.
While the absence of sign epistasis implies the absence of multiple fitness optima, yielding so called \textit{smooth}
landscapes, sign epistasis leads to \textit{rugged} landscapes, which may contain many local fitness optima.

For a long time, the absence of experimental data lead to an exclusively theoretical approach to the study of
fitness landscapes. Over the last years an increasing number of
experimental studies devoted to the analysis of empirical fitness
landscapes and a comparison to theoretical
models have been reported \cite{deVisser2014,Weinreich2006,Poelwijk2007,franke2011,Szendro12,Neidhart13,Weinreich2013}.
Many of these empirical data sets suggest that realistic fitness landscape are not
smooth but show a varying amount epistasis. 
Therefore, some of the effects described above, which assume that
mutations accumulate additively (e.g., the Fisher-Muller effect),
may become less important for the advantage of recombination, even if they are
crucial on smooth landscapes. On the other hand, on a rugged landscape
it becomes more important
for a population to be able to escape from local fitness optima and not only
to find genotypes with larger fitness as fast as possible. It was
shown in \cite{weinreich2005n2,weissman2010,altland2011} that the time a population takes to cross a fitness
valley may have a minimum at a small (but non-zero) recombination rate. However,
when the recombination rate is increased further, the escape time increases
rapidly and even diverges in the limit of infinite population size \cite{Jain2010,Park2011}.
Hence the mechanisms conveying a benefit of recombination 
are in competition with the disadvantageous effect of trapping at local optima.

So far, only a few studies have addressed the problem of recombination
on large fitness landscapes containing sign epistasis
\cite{kondrashov2001,watson2005,devisser2009,bonhoeffer2009,watson2011,moradigaravand2012}
and the criteria that determine whether 
recombination is of advantage on some specific landscape remain unclear. 
Of particular relevance to the present work are the results of a
recent study on recombination in an epistatic fitness model
in which two regimes of evolutionary
dynamics were identified \cite{neher2009,neher2013}: One where recombination is
strong compared to selection, linkage is weak and selection acts mainly on the allele
frequencies, and a weak recombination regime, where linkage is strong and the
dynamics leads to the condensation of the population around particularly fit haplotypes.
In the following, we study the (dis)advantage of recombining populations
compared to non-recombining ones on multidimensional, epistatic fitness landscapes with tunable
ruggedness and strong selection, resulting in strong linkage. Since this corresponds to the weak recombination
regime of \cite{neher2009,neher2013}, a parametrization in terms of linkage disequilibria
and allele frequencies is not suitable and a genotype-based description
is used. 

We find that whether recombination is of advantage or not
depends crucially on the time at which the advantage is evaluated.
Our results are based on simulations of sexual and asexual
Fisher-Wright dynamics on landscapes generated according to the Rough
Mount-Fuji model \cite{aita2000,aita2000a,franke2011,neidhart2014}
(see \textbf{Models and Methods}). In this model the ruggedness of the
landscape can be easily tuned between the additive and fully random
limits, and it has been shown to faithfully reproduce
many statistical features of real fitness landscapes \cite{franke2011,Szendro12,Neidhart13}   
We find numerically that in most cases
recombination is advantageous at intermediate timescales. However, this advantage
is only transitory and non-recombining populations always overtake
the recombining ones in the evolutionary race at long times.
The following sections focus on describing and
explaining the temporal patterns of the relative fitness evolution 
in static landscapes. We then consider evolutionary dynamics on
fitness seascapes \cite{mustonen2009}, i.e.~time dependent fitness landscapes,
to model a population in a changing environment, and show that under this
scenario the advantage of recombination becomes stationary in agreement
with the Red Queen hypothesis.
Although a transitory advantage of recombination has been found and explained previously for
different fitness models (see for example \cite{Feldman1997}), we will argue that the ruggedness of
the landscape introduces a very different origin of this phenomenon.
The same holds true for the associated maintenance of the advantage on time-dependent landscapes
(see \cite{Buerger2000,Otto2009} and references therein).

\section*{Results}
\subsection*{General Phenomenology\label{sec:phenomenology}}

In this article, we are interested in the advantage recombination provides to 
adaptation compared to dynamics consisting only of mutation and selection. Our
model aims at describing haploid populations of facultative sexuals, which are 
common among microorganisms such as bacteria, viruses or rotifers. For this purpose 
we introduce the parameter $r$ as the fraction of the population that
recombines. 
As recombination requires the coexistence of multiple mutant clones to
have an effect, in the following, we choose the population
size $N$ and genome-wide mutation probability $\mu$ such that
$N\mu\geq 1$, which implies that several mutations occur each
generation (see \textbf{Models and methods} for a precise definition
of the model parameters). 
In order to quantify the advantage of recombination, we consider the difference of mean fitness of the two processes,
$\Delta w(t)=\langle w_\mathrm{r}(t)- w_\mathrm{nr}(t)\rangle$, where
$w_\mathrm{r}$ and $w_\mathrm{nr}$ denote the population mean
fitness at time $t$ of a recombining (recombination fraction $r=1$) and a non-recombining ($r=0$) population on the same realization of the
fitness landscape, respectively. Angular brackets
$\langle\ldots\rangle$ denote averaging over (typically a few
thousand) runs of the
dynamics. Each
simulation starts with the whole population placed at the reference genotype $\sigma^\ast$ of a new landscape, created according to the RMF
model with $L=16$ biallelic loci (see \textbf{Models and
  methods}). Since the additive fitness component in the RMF model increases linearly
with distance from $\sigma^\ast$, this corresponds to an initial
condition of low fitness mimicking a poorly adapted organism,
e.g. after a change of environment, which is chosen here in order to
provide a long period of adaptation. Note, however, that the overall
phenomenology remains valid for different choices of
the starting genotype.

As an alternative measure we considered the probability that the recombining population has a higher mean fitness than the non-recombining one, $P_\mathrm{adv}(
t)=\langle\theta(w_\mathrm{r}(t)-w_\mathrm{nr}(t))\rangle$, where
$\theta(x)$ denotes the Heaviside theta function that equals unity for
$x>0$ and zero for $x \leq 0$. 
Another possibility would be to consider a single population into
which a modifier allele is introduced that determines
whether an individual proliferates sexually or asexually \cite{Nei1967}.
In this setup recombining and non-recombining populations are under direct competition. The fraction of
sexuals in a population could then again be used as an indicator for the
(dis)advantage of recombination.
As can be verified in 
fig.~\ref{padv}, both alternatives behave similarly to $\Delta
w(t)$. The semblance to  $P_\mathrm{adv}$ implies that $\Delta w(t)$ is not dominated by a few
exceptional runs of the dynamics. Hence it is an appropriate choice
for an indicator 
to assess whether recombination is typically of advantage or not and we will mainly concentrate on this measure throughout the article.

%%%%%%%%%%%%%%%%%%%%%%%%%%%%%%%%%%%%%%%%%%%%%%%%%%%%%%%%%%%%%%%%%%%%%%%%%
\begin{figure}[htb]
\includegraphics[width=0.8\columnwidth]{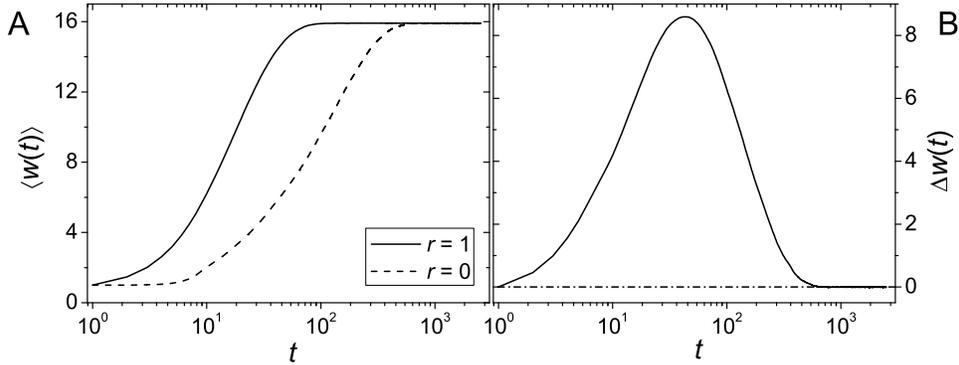}
\caption{\textbf{Advantage of recombination on a smooth fitness
    landscape.} The figure shows time series of (A) mean fitness~$\langle w\rangle$ and (B)
  difference of the mean fitnesses $\Delta w$ for recombining ($r=1$)
  and non-recombining ($r=0$) populations in an additive
  landscape. Time $t$ is measured in discrete generations. Population parameters are $N=1000$, $N\mu=8$, and the
  slope of the additive fitness landscape is $c=1$. Although not visible to the naked eye, $\Delta w$ converges to a value below zero at long times.\label{additive}}
\end{figure}
%%%%%%%%%%%%%%%%%%%%%%%%%%%%%%%%%%%%%%%%%%%%%%%%%%%%%%%%%%%%%%%%%%%%%%%%

Before considering fitness landscapes with sign epistasis, it is instructive to have a first short look at purely additive landscapes. 
As can be seen in fig.~\ref{additive}B, $\Delta w$ increases in time until it reaches a maximum and subsequently declines to a value around, actually slightly below, zero. 
This behavior is easily understood when considering the individual curves for $\langle w_\mathrm{r}\rangle$ and $\langle w_\mathrm{nr}\rangle$ (see fig.~\ref{additive}A). 
For small times $\langle w_\mathrm{r}\rangle$ increases more rapidly than $\langle w_\mathrm{nr} \rangle$ until it converges to a maximal level. 
Subsequently $\langle w_\mathrm{nr}\rangle$ catches up and converges to a
slightly higher value (see e.g.~\cite{nagylaki1993} for an analytical approach
to a similar setting in the weak selection limit). 
It is easy to check that the initial increase of $\langle w_\mathrm{r}\rangle$ and $\langle w_\mathrm{nr}\rangle$ stops when the populations reach the global optimum of the landscape. 
This setup without epistasis has been 
studied in some detail in \cite{kim2005}, where the initial advantage of recombination was attributed to the Fisher-Muller effect. 
The slight disadvantage of recombination at long times reflects the
recombination load, i.e., the fact that recombining populations maintain a
larger amount of genotypic diversity, which means a larger number of
individuals carrying not the fittest but neighboring genotypes.
In the setup with direct competition \cite{Nei1967},
the non-recombining subpopulation would have
a selective advantage at this point which
decreases the amount of recombination. This is known as the reduction principle
\cite{Feldman1986}. Note however that if selection is weak enough, recombining
populations will asymptotically achieve higher mean fitness values than non
recombining ones due to the Hill-Robertson effect\cite{Barton2010}.

%%%%%%%%%%%%%%%%%%%%%%%%%%%%%%%%%%%%%%%%%%%%%%%%%%%%%%%%%%%%%%%%%%%%%%%%%
\begin{figure}[h!]
\includegraphics[width=0.8\columnwidth]{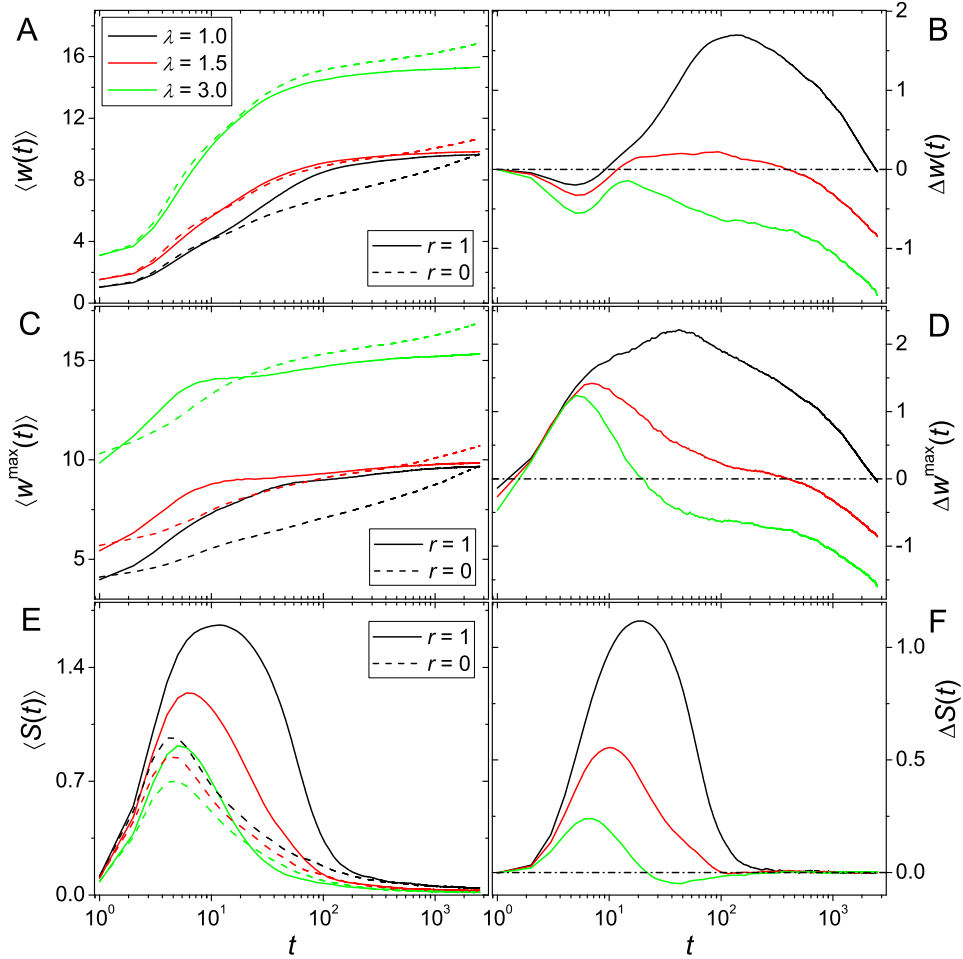}
\caption{\textbf{Dependence of the recombinational advantage on
    fitness landscape ruggedness.} The figure shows time series of different observables for varying ruggedness parameter $\lambda$.
(A,~B) Mean fitness~$\langle w\rangle$.
(C,~D) Fitness of the fittest individual~$\langle w^\mathrm{max}\rangle$.
(E,~F) Entropy~$\langle S\rangle$ of the genotype frequency
distribution as a measure for genetic diversity.
Left columns shows quantities for populations with $r=0$ and $r=1$ separately, the right column
shows the difference. Parameters are $N=2000$, $N\mu=4$, and $c=1$.\label{deponlambda}}
\end{figure}
%%%%%%%%%%%%%%%%%%%%%%%%%%%%%%%%%%%%%%%%%%%%%%%%%%%%%%%%%%%%%%%%%%%%%%%%

Let us now turn to rugged landscapes. In fig.~\ref{deponlambda}B we
show results for $\Delta w(t)$ corresponding to three different
choices of $\lambda$, the parameter of the
exponential random variables describing the random component of the
RMF fitness landscape (see \textbf{Models and Methods}); further
examples are shown in figs.~\ref{deponNmu}A,B and \ref{deponN}A,B. 
Now $\Delta w$ becomes clearly negative at long enough times, and not (close to) zero as before.
This means that non-recombining populations always overtake recombining ones.
The origin of this long term disadvantage can again be better
understood by considering $ w_\mathrm{r}$ and $w_\mathrm{nr}$ separately (see fig.~\ref{deponlambda}A).
A crucial observation is that the increase of fitness of recombining populations levels off very fast at some point, while for non-recombining populations it increases slowly but more steadily. 
Unlike in the case without epistasis, the leveling off for recombining
populations is not because the global optimum has been reached, and a higher
genetic variance induces a recombination load.
As we will show later on, the explanation is that populations are trapped for extremely long times at local maxima.
We will argue that such trappings are the more likely, and the trapping times the longer, the larger the recombination rate.
Therefore, in rugged landscapes, large recombination rates are always disadvantageous at long times.
Nonetheless, figs.~\ref{deponlambda}B, \ref{deponNmu}A,B and \ref{deponN}A,B also clearly show a 
transitory regime where recombination can be advantageous, depending
on the parameters controlling the ruggedness of the landscape and the
dynamics.

The temporal pattern observed for $\Delta w$ appears to be qualitatively the same on any rugged landscape, 
including models other than RMF (such as the NK-model \cite{kauffman1989}, see fig.~\ref{lk_kdep}).
At very short times non-recombining populations are more efficient in
adapting, causing an initial decrease of the $\Delta w$.
Shortly after $\Delta w$ increases again, showing that now recombination is more efficient.
Finally, this period of fast fitness increase of the recombining
population comes to an end and the non-recombining populations again
display a faster fitness increase.
Whether the intermediate period of increased effectiveness of
recombining populations is long enough to catch up and temporarily 
overtake the non-recombining populations depends on the systems parameters.

Albeit of transitory nature, this advantage of recombination at intermediate timescales may be essential for a population that is transferred to an environment where it is initially poorly adapted.
Although the effect is transient, the additional gain in adaptive effectiveness could determine if the population adapts in time before being displaced by other organisms.
Further, if the period of positive $\Delta w$ is long enough, a non-recombining populations might be displaced by the recombining variant before it can make use of its long time advantage.
The remainder of this article is devoted to explaining the processes
responsible for the observed temporal pattern of $\Delta w$.

\subsection*{Short time advantage of non-recombining populations\label{sec:small_time}}
As can be verified in figs.~\ref{deponlambda}B, \ref{deponNmu}A,B there is a short period at small times where $\Delta w$ decreases,
implying an advantage of non-recombining populations (see also \ref{deponN}A,B in the supplementary material).
We will argue that the initial decrease of $\Delta w$ is a consequence of the monomorphic initial condition,
or, more precisely, of the property of recombination to create larger
diversity which is also at the heart of the concept of recombination load. 
Non-recombining populations rapidly select some fit neighboring state
and reside there almost monomorphically until a new fitter mutant is created. 
The recombining populations also mostly occupy such fit neighbors, but
in addition contain a considerable fraction of other genotypes that
are often less fit.
This lowers the average fitness but is not necessarily a
disadvantage for further adaptation, as the diversity can augment the
evolvability of the population.

To show that recombining populations do indeed create a larger diversity, we have calculated the Shannon entropy $S(t)$ of the population distribution, which is defined as 
\begin{equation}
S(t)=-\sum_{\sigma} f_\sigma(t) \ln{[f_\sigma(t)]}. 
\end{equation}
Here the sum runs over all genotypes, and $f_\sigma(t)$ is the fraction of the population with genotype $\sigma$ at time $t$. 
The more homogeneously the population is distributed among the different states, the larger is the value of $S$, while for a monomorphic population $S=0$. 
In the following we consider $\langle S(t)\rangle$,
where the brackets again denote averaging over ensembles of simulation runs,
each in a different realization of the landscape. An alternative measure for the
diversity is the additive genetic variance $V^\mathrm{a}$, which is the variance of the part of
the fitness which is inherited. In our setup, this is just the variance of the distance to
the reference sequence in the population, $V^\mathrm{a} = \text{Var}[d(\sigma,
\sigma^\ast)]$ (see \textbf{Model and Methods} for the definitions of $\sigma^*$ and
$d(\cdot, \cdot)$). As can be seen in fig.~\ref{Sandvarn}, $\langle
V^\mathrm{a} \rangle$ and $\langle S\rangle$ behave
very similarly and we will therefore, in the following, only discuss 
the entropy.

For recombining as well as non-recombining populations $\langle S(t) \rangle$ initially increases at short times but then decreases again to a very low level (cf.\ fig.~\ref{deponlambda}E).
In the initial period $\langle S(t) \rangle$ is always larger for recombining than for non-recombining populations.
Further, the initial increase of $\langle S(t) \rangle$ is faster and also more prolonged.
This clearly shows the ability of recombination to build up a larger
amount of diversity, supporting our claim that the initial decrease of
$\Delta w$ is due to recombining populations distributing 
among a large number of states many of which are probably not very fit.   

However, there are at least two other mechanisms that could contribute
to the decrease: First, non-recombining populations might be more
efficient in finding particularly fit mutants, and second, 
non-recombining populations might be more effective in selecting fit mutants once they have been created.

The first claim is easy to disprove. 
In fig.~\ref{deponlambda}D we plot $\Delta w^\mathrm{max}(t)=\langle w^\mathrm{max}_\mathrm{r}(t) - w^\mathrm{max}_\mathrm{nr}(t)\rangle$,
where $w^\mathrm{max}_\mathrm{r}(t)$ and $w^\mathrm{max}_\mathrm{nr}(t)$ denote the fitness of the fittest observed mutants at time $t$ in recombining and non-recombining populations, respectively.
The temporal pattern of $\Delta w^\mathrm{max}$ is similar to the one
of $\Delta w$, but there is no dip at short times (except perhaps a
very small disadvantage at the first two time steps). 
This shows that even at short times recombining populations are better in finding particularly fit mutants than non-recombining ones.
Note also that these particularly fit mutants have a high probability
to take over the population due to selection and therefore determine the future evolution. 
This might be the reason why the temporal patterns of $\langle
w^\mathrm{max}\rangle$'s and $\Delta w^\mathrm{max}$ are similar in
shape to the ones of the $\langle w\rangle$'s and $\Delta w$, but with all features shifted to shorter times (see figs.~\ref{deponlambda}A,B,C,D).
In fact, in order to evaluate which dynamics performs better, it might
be more appropriate to consider $\Delta w^\mathrm{max}$ instead of
$\Delta w$. 

The second claim, that the initial decline may reflect non-recombining populations being more effective in selecting fit mutants, is not that easy to reject. In fact, the very small negative values of $\Delta w^\mathrm{max}$ at the first two time steps might be due to this reason. In our setup recombination acts as an additional source of genetic drift, which augments the probability of extinction of newly found beneficial mutants. As a consequence beneficial mutants fix less effectively. However, for large populations this effect should be small.

\subsection*{Period of advantage of recombination\label{sec:medium_time}}
A large part of the literature about sexual reproduction deals with the question by which mechanisms recombination can speed up evolution, two of the most prominent ones being the Fisher-Muller and the Weismann effect.
Despite the considerable amount of dedicated research, it has proven rather difficult to assess the impact of the different mechanisms, especially when considering rugged fitness landscapes for large genomes. 
In this article, we are mainly interested in understanding why these mechanisms seem to fail at long times. 
Therefore, we restrict ourselves to the observation that the standard
mechanisms can be efficient at intermediate times, as shown by the increase of $\Delta w$.
This regime depends in rather complex ways on the parameters controlling the landscape ruggedness and the dynamics. 

The Fisher-Muller effect consists in the advantage gained by combining mutations present in the population to obtain new, fitter recombinants, 
allowing the genotype space to be explored in large jumps. 
This mechanism should be efficient on a RMF landscape as on such a landscape the fitness increases on average the more mutations are accumulated with respect to the initial genotype $\sigma^\ast$. 
Of course, recombination can also lead to the loss of mutations.
But as the mutants containing many mutations are expected to be
particularly fit, they have a large probability to take over the
population due to selection, whereas
less fit recombinants are rapidly purged and do not affect the future 
development of the population.

In the presence of epistasis, one could naively expect an even larger advantage of recombination,
because the neighboring states of highly populated genotypes are
typically of low fitness, whereas the large jumps due to recombination
may lead to states much further uphill the landscape
that have a high probability to be more fit.
As we will see later, this intuition is largely wrong,
since recombination enhances the trapping at local maxima.
Nonetheless, before the population falls into such a local maximum, this mechanism may still be efficient.

\subsubsection*{Effect of landscape ruggedness}
In fig.~\ref{deponlambda}B we plot $\Delta w$ for three different
choices of the ruggedness parameter $\lambda$. 
While the advantage of recombination in the intermediate regime decreases monotonically with increasing ruggedness, the effect of $\lambda$ on the individual mean fitnesses $\langle w(t)\rangle$ is less clear. 
Figure~\ref{deponlambda}A seems to suggest that both $\langle
w_\mathrm{r}(t) \rangle$ and $\langle w_\mathrm{nr}(t) \rangle$
increase with increasing $\lambda$, but this behavior is confounded by
the fact that the random component of the landscape has mean
$\lambda$, and therefore the mean fitness level shifts with increasing
$\lambda$ (see \textbf{Models and methods}). There are also cases
where higher mean fitness values are reached for smaller $\lambda$,
but it remains true that $\Delta w$ decreases with increasing $\lambda$.

A possible explanation for the decrease of the adaptive advantage of
recombining populations with increasing landscape ruggedness lies in
the link to population diversity, which is a prerequisite for both
the Fisher-Muller effect and the Weismann effect.
In rugged landscapes diversity is suppressed, because selection
focuses the population onto high-lying fitness peaks and ridges
separated by valleys of much lower fitness.
On smoother landscapes, on the other hand, many mutants have comparable fitness values and will therefore coexist for longer times. 
This effect can be observed in recombining as well as in
non-recombining populations, as illustrated by the behavior of the
corresponding mean entropies fig.~\ref{deponlambda}E. Recombining
populations create a larger diversity than non-recombining ones, such
that the entropy difference $\Delta S=\langle S_\mathrm{r} -
S_\mathrm{nr}\rangle$ is positive at small times and increases until
it starts to decrease at some time that shortly anticipates the
decrease of $\Delta w$ (compare figs.~\ref{deponlambda}B and F). 
Importantly, fig.~\ref{deponlambda}F also shows that $\Delta S$ decreases with increasing $\lambda$. 

%%%%%%%%%%%%%%%%%%%%%%%%%%%%%%%%%%%%%%%%%%%%%%%%%%%%%%%%%%%%%%%%%%%%%%%%%
\begin{figure}[htb]
\includegraphics[width=0.8\columnwidth]{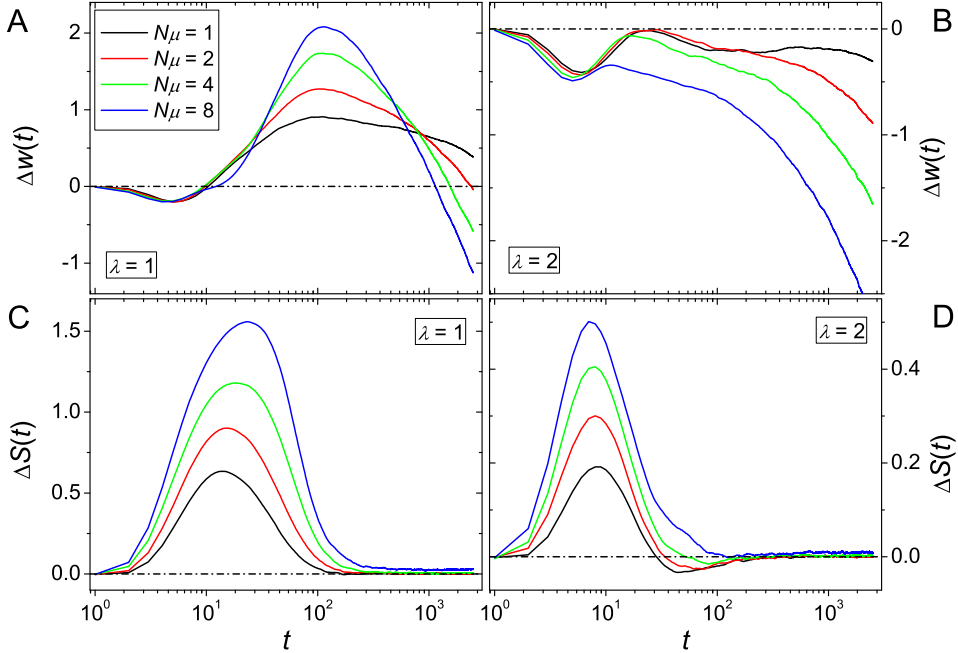}
\caption{\textbf{Dependence of the recombinational advantage on
    mutation supply.} The figure shows time series of mean fitness difference $\Delta w$ (A,~B) and entropy 
difference $\Delta S$ (C,~D) for varying mutation rate $\mu$ and two different
values of the ruggedness parameter $\lambda$. The population size
is $N=1000$ and $c=1$.\label{deponNmu}}
\end{figure}
%%%%%%%%%%%%%%%%%%%%%%%%%%%%%%%%%%%%%%%%%%%%%%%%%%%%%%%%%%%%%%%%%%%%%%%%

\subsubsection*{Effect of mutation supply and population size}
A simple relation between population diversity and rate of adaptation
is suggested by Fisher's fundamental theorem, which states that the
mean fitness increase is proportional to the fitness variance
\cite{fisher1930}. In the following we show that this argument is not generally valid on rugged fitness landscapes. 
For this purpose we manipulate the population diversity through the
mutation supply rate $N \mu$, the number of mutants created in each generation.
With increasing $N\mu$ the population diversity quantified by the
entropy $S$  increases for recombining as well as for non-recombining
populations (not shown).
The effect is again larger for recombining populations, as can be
verified in figs.~\ref{deponNmu}C,D. 
However, whereas for the smoother landscape ($\lambda=1$) the increase
in diversity is accompanied by an increase in $\Delta w$, 
the opposite is the case for the more rugged landscape ($\lambda=2$) (figs.~\ref{deponNmu}A and B). 

The reason for this negative influence of increased diversity on the advantage of recombination on very rugged fitness landscapes may be the following:
The large diversity at initial times causes the coexistence of many
different mutants, but only the most fit mutants have the chance to get selected. 
On very rugged landscapes, such mutants have a relatively high probability to be local optima. 
As will be shown in the next section, once almost the entire
population resides on such a local optimum, the recombining population essentially stops adapting. 
On less rugged landscapes, on the other hand,
the probability to find a local optimum is smaller 
and hence the population can derive an extended benefit from the larger diversity.

A similar but less pronounced change in population diversity also mediates the effect of
the population size on the advantage of recombination, when $N$ is
varied at fixed mutation supply rate $N \mu$.
As $N$ is increased, the strength of selection is enhanced while
genetic drift decreases. In addition, we claim that the diversifying
effect of mutation is reduced when $N$ increases, leading to a
decrease of the entropy $S$. In the presence of recombination, this
decrease is further enhanced, 
leading to a decrease of $\Delta S$ (see figs.~\ref{deponN}C,D in the supplementary material).
This behavior can be understood from the following argument: 
In each time step, $m=N\mu$ mutants are created on
average. The frequencies $\tilde f_{j} = \frac {m_j}{N}$
corresponding to the different genotypes of newly produced mutants 
are inversely proportional to $N$. 
Therefore, the larger $N$, the smaller the fraction of the population
exploring new genotypes.
This, in turn, leads to a lower diversity for both recombining and
non-recombining populations.
But as the frequencies of mutants decrease, so does the frequency of new
individuals generated by recombination,
Given two new states populated with frequencies $\tilde f_i$ and $\tilde
f_j$, on average, a fraction $2r\tilde f_i\tilde f_j \propto N^{-2}$
of the population will be provided by the recombination of their genotypes.
Thus, for increasing $N$, the absolute number of offspring of such
pairings is diminished, suppressing the exploration of additional
unpopulated sequences by recombination. This causes the decrease of
$\Delta S$.
Just as for the case where $N\mu$ was varied (fig.~\ref{deponNmu}), the
decrease in diversity leads to a somewhat
smaller adaptive advantage of recombination in relatively smooth landscapes,
while the opposite is true in more rugged landscapes
(see figs.~\ref{deponN}A,B in the supplementary material).

\subsubsection*{Infinite populations}
A question that naturally arises here is whether the transient advantage of recombination still exists in the limit $N\rightarrow\infty$.
A priori this is not at all obvious. If the only advantage of recombination would lie in the exploration of formerly undiscovered genotypes,
it would be irrelevant in the infinite $N$ limit, as in this limit all genotypes are populated after the first mutational step.
On the other hand it is also not clear if trapping at local fitness maxima will still occur if all, even distant, maxima are populated.

In fact, trapping has been observed in the infinite population
dynamics on a specific, rugged empirical fitness landscape \cite{devisser2009}. To
clarify whether this is a common phenomenon, we performed simulations in the infinite $N$ limit on RMF landscapes.
The algorithm allows for an infinite number of multiple mutations in a
single time step and is described in \textbf{Models and methods}. 

%%%%%%%%%%%%%%%%%%%%%%%%%%%%%%%%%%%%%%%%%%%%%%%%%%%%%%%%%%%%%%%%%%%%%%%%
\begin{figure}[htb]
\includegraphics[width = 0.8\columnwidth]{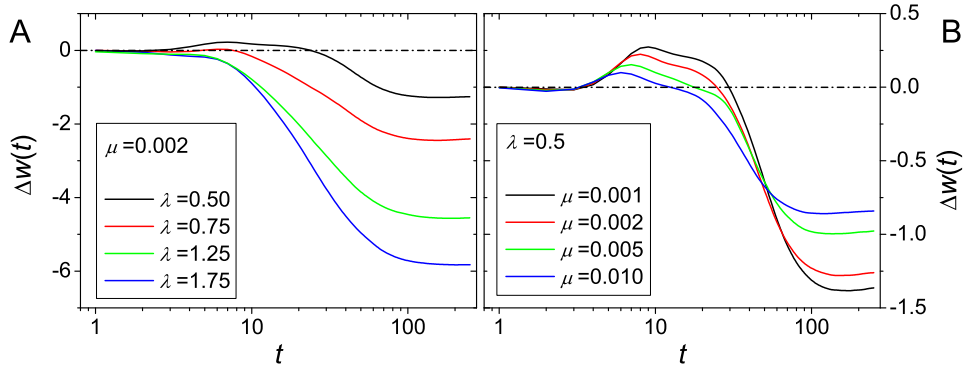}
\caption{\textbf{Advantage of recombination for infinite populations.}
  The figures show the results of simulations in the infinite population size limit. (A) shows a comparison for different choices  of $\lambda$ and (B) a comparison for different mutation
rates $\mu$.
}
\label{fig:Minf}
\end{figure}
%%%%%%%%%%%%%%%%%%%%%%%%%%%%%%%%%%%%%%%%%%%%%%%%%%%%%%%%%%%%%%%%%%%%%%%555

Figure \ref{fig:Minf}A shows the results of simulations with $L=8$ and various choices of the landscapes ruggedness. The general shape of
$\Delta w$ is similar to that observed in finite populations. At least
for less rugged landscapes, recombination is still
beneficial on intermediate time scales. 
Hence we conclude that recombination not only provides an advantage by
exploring a larger fraction of the landscape but may also speed up the
adaptation process. With increasing ruggedness, trapping events happen
more often and earlier, eventually eliminating the transient advantage
of recombination. 

A comparison of different mutation rates is shown in fig. \ref{fig:Minf}B. Larger values of $\mu$ lead to
a faster fitness increase for mutation-only processes, decreasing the
advantage of recombination at intermediate times. On the other hand,
the mean fitness reached by the non-recombining populations in the
long time limit (where $\Delta w < 0$) is higher for small values
of $\mu$. This is due to the mutational load: the larger mutation rate leads to a larger fraction of the population spreading over
the landscape, even after the global maximum has already fixed.

\subsection*{Long time disadvantage of recombination\label{sec:long_time}}

Although the mechanisms leading to an advantage of recombination work
efficiently at intermediate times, we have already seen that this advantage is only transitory. 
In the following, this finding will be explained on the basis of known
results concerning the dynamics of evolution with recombination in a two locus system. 
For that purpose we recall some of these results that will prove
essential to understanding the dynamics in high-dimensional fitness
landscapes, see \cite{weinreich2005n2,weissman2009,weissman2010,altland2011,Jain2010,Park2011} for details.

Consider a population that reached a local fitness maximum. To proceed and reach states with higher fitness, a fitness valley has to be crossed. 
Populations that do not recombine can do so in two ways. 
The first way consists in first  fixing the population at a valley
genotype and subsequently  fixing it at some state fitter than the initial
state. Here, ``fixation'' is meant in an approximative way, in the sense that
the population becomes strongly concentrated on some genotype.
The second way to cross the valley is to produce only a small population of mutants in the valley, which can then further mutate, giving rise to new mutants at a larger distance from the initial genotype. 
These initially few multiple mutants can, if they are fit enough, take over the population. 
Both modes of valley crossing are strongly suppressed with increasing population size $N$, but for large enough $N$ the second one dominates \cite{weinreich2005n2,weissman2009}. 

Recombination opens up a third path. Mutants can be produced at
various states in the valley and these mutants can then recombine to
produce fitter mutants at larger distance from the initial genotype. 
But apart from providing a new path to escape from a local maximum, recombination also yields a mechanism that can make the escape even more difficult. 
Suppose that a few mutants have been produced on a state with larger fitness than the initial one. 
Since nearly the whole population is still located on the initial
(peak) state, recombination replaces these fit mutants by unfit valley
genotypes with high probability, and selection will then resort them to the initial point. 

Whether recombination is of advantage or disadvantage for such escapes
depends strongly on the selective benefit of the target state and the
selective disadvantage of the valley states
in relation to the fraction of recombining individuals $r$. 
Most noteworthy, if $r$ is larger than some
critical value $r^*$ which depends on the selection coefficients, the
escape time in the two-locus landscape increases exponentially with $N$, 
at $r\approx r^*$ as $\ln{t_\mathrm{esc}}\sim N(r-r^*)^{3/2}$ \cite{altland2011}.
In contrast, the escape times for non-recombining populations increase
only algebraically in $N$\cite{weinreich2005n2,weissman2009}. 
In our multidimensional landscape model, each possible escape path
from a local peak involves different fitnesses and hence a different value of $r^*$.
For large $N$ it follows 
that the escape times will become extremely large compared to those of non-recombining populations, 
if a major part of the population reaches a local maximum at which all the $r^*$'s are smaller than $r$.
The recombining populations are trapped.
As a consequence, the advantage that the recombining populations could build up by means of the Fisher-Muller 
and the Weismann effect is lost, 
while the non-recombining ones can slowly catch up and overtake. In
the limit of infinite population size these local maxima in the genotypic fitness 
landscape can induce multiple equilibrium states, in the sense that, depending on the initial condition, the 
population remains centered around a suboptimal fitness peak for all times. Again, this phenomenon occurs
when the recombination rate exceeds a certain threshold and is absent without recombination 
\cite{Jain2010,Park2011,Feldman1971,Rutschman1994}.  

%%%%%%%%%%%%%%%%%%%%%%%%%%%%%%%%%%%%%%%%%%%%%%%%%%%%%%%%%%%%%%%%%%%%%%%%%
\begin{figure}[htb]
\includegraphics[width=0.6\columnwidth]{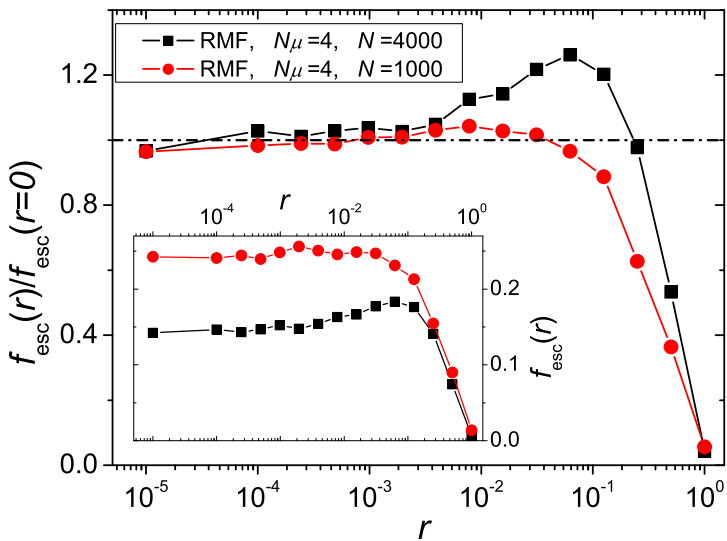}
\caption{\textbf{Escape from local maxima.} Fraction of escape events
$f_\mathrm{esc}$ on RMF landscapes with $\lambda = c = 1$
in dependence on the recombination rate $r$ and normalized to the value at $r=0$.
The inset shows the values without normalization. For suitable choices
of the population parameters $f_\mathrm{esc}(r=1)/f_\mathrm{esc}(r=0)$ has a
maximum. \label{escape}}
\end{figure}
%%%%%%%%%%%%%%%%%%%%%%%%%%%%%%%%%%%%%%%%%%%%%%%%%%%%%%%%%%%%%%%%%%%%%%%%

To show that the trapping scenario known from the two locus system is also effective in high-dimensional landscapes, 
we have measured the fraction of escape events over the number of trapping events, $f_\mathrm{esc}=n_\mathrm{esc}/n_\mathrm{trap}$, 
observed up to run time $t=500$ as a function of the recombination
rate $r$. 
Note that, because each escape is preceded by a trapping event,
$f_\mathrm{esc}\rightarrow 1$ for very long times. 
However, in the cases studied here, escapes are very rare, and $f_\mathrm{esc}$ is still a useful measure.  
In the simulations we considered a population as trapped if $70\%$ of the individuals share the genome corresponding to the local maximum and an escape event is registered if, for a population that was marked as trapped, the fraction of mutants on the local maximum falls below $50\%$. 

As expected, the number of escapes drops rapidly when $r$
approaches unity (fig.~\ref{escape}). 
As $r$ decreases, strong trapping, that only occurs for local maxima for which all the $r^*$'s are smaller than $r$, gets less likely and the number of escapes converges to that expected in the absence of recombination.      
The decrease of $f_\mathrm{esc}$ when increasing $N$ (see inset of
fig.~\ref{escape}) is not surprising since both the escape under the
influence of only mutation, as well the escape in the presence of recombination, 
are suppressed when increasing $N$ \cite{weinreich2005n2,weissman2010,weissman2009}.
Interestingly, for sufficiently large $N$, there exists an intermediate regime of $r$ values, where escapes are more frequent with, than without recombination. 
This can again be explained by the results of the two locus model. 
There, it was shown that, if the number of individuals at the valley
states is dominated by fluctuations, the escape time has a minimum \cite{weissman2010,altland2011}. 
We claim that this minimum in the escape times is the cause of the maximum in the escape fraction, which is
supported by the following observation. 
The number of mutants that are not located at the most populated genotype when the latter is a local optimum
is a proxy for the number of valley mutants.  
Its coefficient of variation increases with $N$ for the parameters used in fig.~\ref{escape} (see
fig.~\ref{cvar}). Thus the effect of fluctuations increases with $N$
in this case, which explains why an `optimal' value of $r$ appears for
$N=4000$ but not for $N=1000$.

%%%%%%%%%%%%%%%%%%%%%%%%%%%%%%%%%%%%%%%%%%%%%%%%%%%%%%%%%%%%%%%%%%%%%%%%%
\begin{figure}[htb]
\includegraphics[width=0.8\columnwidth]{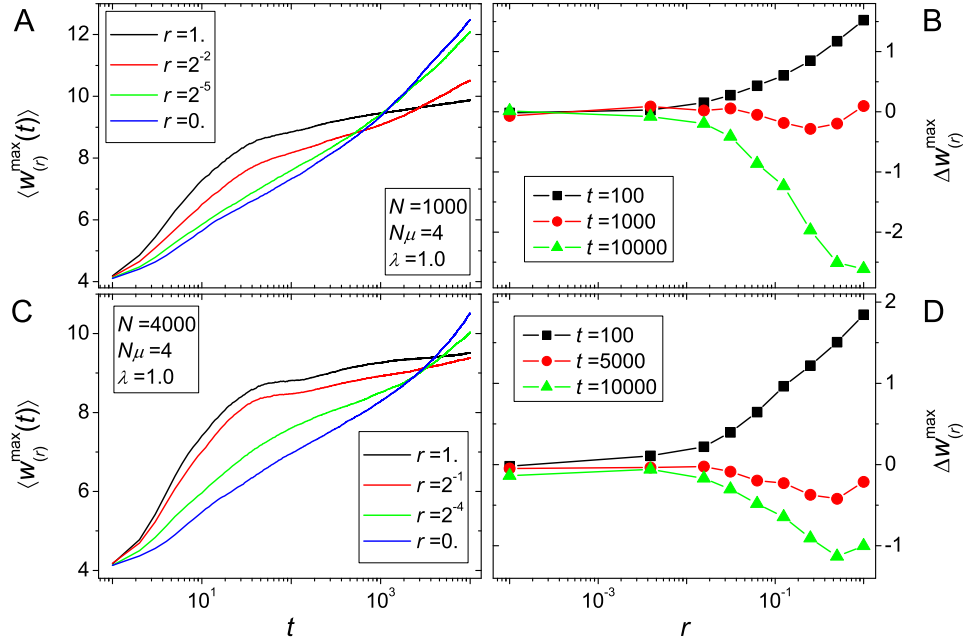}
\caption{\textbf{Recombinational advantage as a function of
    recombination rate and observation time.} (A) and (C): Fitness of the fittest genotype $w^\mathrm{max}_{(r)}$ vs.\ time for various choices of the recombination rate $r$. Parameters correspond to those presented in fig.~\ref{escape}. For short times $w^\mathrm{max}_{(r)}$ grows more rapidly for large $r$ but at long times $w^\mathrm{max}_{(r)}$ grows faster for small $r$. At intermediate times the acquired fitness depends non-monotonically on $r$. (B) and (D): $\Delta w^\mathrm{max}_{(r)}$ vs.\ $r$ for various choices of the evaluation time $t$. Whether high recombination rates are advantageous depends on evaluation time. At intermediate times, the dependence of recombinational advantage on $r$ is non-monotonic. Note that for both cases shown, $\Delta w^\mathrm{max}_{(r)}$ has a minimum for intermediate recombination rates at intermediate times.\label{evaladv}}
\end{figure}
%%%%%%%%%%%%%%%%%%%%%%%%%%%%%%%%%%%%%%%%%%%%%%%%%%%%%%%%%%%%%%%%%%%%%%%%

It is however not clear whether the minimum in the average escape time is particularly relevant for the overall
advantage of recombination. The tradeoff between long evolution times before strong trapping, as found for small $r$,
and high rates of adaptation at early times, which occur at large $r$, may be more relevant. A further effect that may be important at this point is a maximum of the fitness variance at intermediate $r$ that has been found in \cite{Turelli1990} for the case of weak selection. A priori, it is difficult to predict
whether this tradeoff leads to a maximal or a minimal advantage of recombination at some intermediate value of $r$.
To assess these questions, we show in figs.~\ref{evaladv}A,C the
temporal evolution of the mean fitness of the fittest genotype $w_{(r)}^\mathrm{max}$ for
different choices of the recombination rates, and in figs.~\ref{evaladv}B,D the value of $\Delta
w_{(r)}^\mathrm{max}=w_{r}^\mathrm{max}-w_{nr}^\mathrm{max}$ as a function of the recombination rates evaluated at a few chosen time points.
For the same choices of parameters as in fig.~\ref{escape}, we do not observe any maximum of $\Delta w_{(r)}^\mathrm{max}$ with respect to $r$ at any time. This indicates that the escape 
rates, which showed a maximum for intermediate $r$, may be of minor importance, at least for the parameter values studied here.

Furthermore, fig.~\ref{evaladv} clearly shows that whether recombination is advantageous or not depends crucially on the
time at which the advantage is evaluated. For short times, at which even for high recombination rates populations
did not yet get trapped, recombination is always of advantage, while
for very long times high recombination rates are always
disadvantageous. At intermediate times we observe a short time window
for which the $r$-dependence of $\Delta w_{(r)}^\mathrm{max}$ is
non-monotonic. For the parameters chosen in figs.~\ref{evaladv}B,D, we
observe a minimum of $\Delta w_{(r)}^\mathrm{max}$ with respect to
$r$, but for other parameter choices an optimal recombination rate can 
exist.

\subsection*{Dependence on the number of loci \label{sec:genome_size}}

So far we considered landscapes with a fixed number of 16 binary loci.
In this section, we are going to discuss the dependence of the dynamics on the number of loci $L$.
As we argued above, the existence of local maxima is crucial for the population 
dynamics. By increasing $L$, the probability of a given genotype being a local
maximum, i.e., the density of local maxima, decreases. For
$\lambda\to\infty$ the density is known to be $\frac{1}{L+1}$ \cite{kaufflev}.
In general the density of maxima in the RMF model depends also on the
distance $d$ to the reference sequence, but it is always a decreasing
function of $L$ \cite{neidhart2014}.
Therefore, both recombining and non-recombining populations adapt
faster for larger $L$ and, as we will see, the recombining ones benefit more from
the thinning of local optima. However, once a recombining population is trapped,
the non-recombining populations exclusively benefit from the lower
trapping probability and can continue their pursuit even faster.

%%%%%%%%%%%%%%%%%%%%%%%%%%%%%%%%%%%%%%%%%%%%%%%%%%%%%%%%%%%%%%%%%%%%%%%%
\begin{figure}[htb]
\includegraphics[width=0.8\columnwidth]{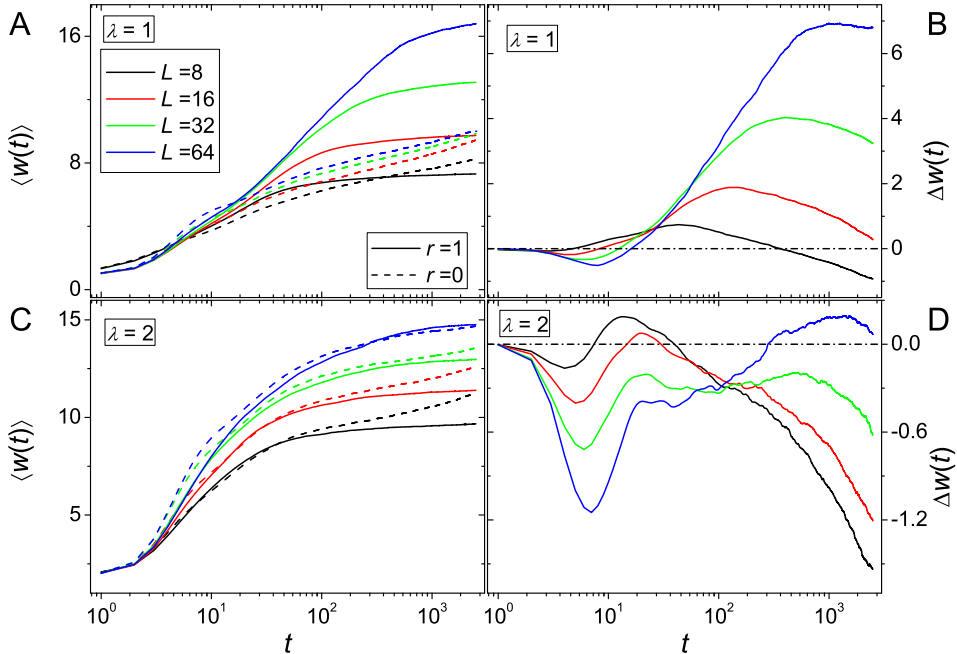}
\caption{\textbf{Recombinational advantage as a function of the number
    of loci.} The figure shows time series of $\langle w \rangle$ and $\Delta
    w$, respectively, for different values of the number of loci $L$. For
    $\lambda = 1$ see (A) and (B), for $\lambda = 2$ see (C) and (D).
The population size is $N=2000$ and the mutation rate is $\mu=0.002$.
\label{fig:Ldep1u2}}
\end{figure}
%%%%%%%%%%%%%%%%%%%%%%%%%%%%%%%%%%%%%%%%%%%%%%%%%%%%%%%%%%%%%%%%%%%%%%%%

Figure~\ref{fig:Ldep1u2} shows fitness time series for different
values $L$. For the smaller ruggedness parameter $\lambda=1$, the
effects that we described before are qualitatively unaltered and
are just amplified due to the increase of $L$.
The initial decline in $\Delta w$ becomes more pronounced, since the ability of
recombining populations to create diversity is enhanced if more neighboring
genotypes are available. 
The intermediate time regime where recombination is advantageous is
prolonged because the probability of being trapped decreases for recombining as well as
non-recombining populations. 
The rate of fitness increase for the recombining populations increases
strongly with the number of loci, 
in accordance with known results for additive fitness landscapes
\cite{kim2005,MaynardSmith1971,Park2013}. 
Hence the advantage of recombination becomes larger and lasts longer (fig.~\ref{fig:Ldep1u2}B).

However, when the ruggedness parameter is increased to $\lambda=2$,
the behavior becomes more complicated.
As can be seen in fig.~\ref{fig:Ldep1u2}C, sexual and asexual populations
derive approximately the same benefit from the increased number of loci.
However, the behavior of the fitness difference changes drastically.
Unlike before, for larger $L$ a second maximum appears in $\Delta w(t)$,
while the first maximum becomes smaller as $L$ increases (cf.\
fig.~\ref{fig:Ldep1u2}D). We will discuss this phenomenon in detail in
the next section. Note further that there is an advantage of
recombination $(\Delta w > 0)$ at
the second maximum when $L$ is sufficiently large. We believe that this remains
true for larger ruggedness parameters, but verifying it by
simulations is not feasible because it would require to increase $L$ to even larger values.

At this point, it is appropriate to ask whether the phenomenology described in the previous sections will remain
valid for realistically large numbers of loci, ranging in the
thousands for genes or in the millions to billions for nucleotide sites. This is not
immediately obvious, because the mechanism responsible for the
phenomenology are the trappings at local maxima and, as we have argued
before, their density decreases when increasing $L$. We will argue
that the answer to this question depends on the ruggedness of the
landscape, as quantified for the RMF model by the ratio between the ruggedness parameter
$\lambda$ and the additive slope parameter $c$ (see \textbf{Models and
  methods}). It is instructive to discuss two extreme cases. 

When $\lambda$ is very small compared to $c$ the landscape is almost
additive and the population most likely reaches the global
optimum. Therefore the distance a population travels before getting
trapped should grow linearly with $L$. 
In fig.~\ref{fig:Ldeptrapn} we show the distance $d_\mathrm{trap}$ of
the genotype at which the population gets trapped for the first time
from $\sigma^\ast$, where trappings are defined as in the last
section.  
For small $\lambda$ the $L$-dependence of $d_\mathrm{trap}$ is well
compatible with a linear relation (fig.~\ref{fig:Ldeptrapn}A in the supplementary material). Therefore, in this regime, when $L$ becomes
large, trapping at local maxima will be very rare and should only play
a minor role in the dynamics, thus invalidating the phenomenology
described earlier.

At the other extreme, when $\lambda$ is large compared to $c$, the
population is likely to end up in a local optimum. Although we do not
have a precise prediction for how far such a population travels before
getting stuck, we can refer to the behavior in the well-studied strong 
selection weak mutation (SSWM) regime, which is characterized by the condition
$N \mu \ll 1$ \cite{gillespie1983,orr2003}. In the SSWM regime the
population is almost monomorphic and evolves as a single entity
performing an uphill \textit{adaptive walk} on the landscape. In the
case of a completely random landscape ($c=0$), the distance a
population travels before it gets stuck grows logarithmically with $L$
\cite{gillespie1983,flyvbjerg1992,orr2003,krugneid,jain2011}.
This result remains valid in the RMF model for small $c$ 
\cite{neidhart2014}. 
Although our dynamics takes place far from the
SSWM regime, the logarithmic prediction is compatible with our
numerical findings for highly rugged landscapes (fig.~\ref{fig:Ldeptrapn}B in the supplementary material). 
If this relation prevails for extremely large $L$, we expect the 
trapping at local maxima to be relevant for realistically large
genomes and the mechanism leading to a disadvantage of recombination
at long times to remain valid.

\subsection*{Time scales and dynamic regimes \label{sec:timescales}}
Having explained the transient nature of the advantage of recombination in qualitative terms, 
the next step would be to estimate the time $t_0$ up to
which recombination is, on average, of advantage, the time
$t_\mathrm{max}$ at which recombination is, again on average, of
largest advantage,
and the corresponding maximal advantage $\Delta w(t_\mathrm{max})$. 
Let $\ell_\mathrm{rec}$ denote the mean distance a recombining
population can travel before it gets stuck in a local maximum, and
$d_\mathrm{mut}(t)$ 
the mean distance at time $t$ from the initial genotype $\sigma^\ast$ of a non-recombining population.
Then $t_0$ can be estimated from
% \begin{equation}
$\ell_\mathrm{rec} = d_\mathrm{mut}(t_0)$.
% \end{equation}
Unfortunately, we do not have analytic expressions for any of these
two quantities and it seems improbable that they should be easy to
obtain. The distance
$\ell_\mathrm{rec}$ is similar to the length of adaptive walks which has been calculated for non-recombining populations 
in the SSWM approximation \cite{gillespie1983,flyvbjerg1992,orr1,orr2003,joyceorr,krugneid,jain2011}. 
This approximation is clearly not applicable for recombining populations, as recombination requires polymorphism. 
Expressions for $d_\mathrm{mut}(t)$ have been derived for populations
evolving beyond the SSWM regime ($N\mu \gg 1$), but these works are
largely restricted to non-epistatic landscapes (for a review see
\cite{Park2010}). The only exception known to us \cite{Park2008} employs an infinite
sites limit, which eliminates the possibility of trapping at local
fitness maxima.   
 
As we argued before, the probability of trapping depends on the
mean density $\rho$ of local maxima.
It therefore seems reasonable to expect that the maximal acquired recombination advantage $\Delta w(t_\mathrm{max})$ should be the larger, the smaller $\rho$. 
A priori, it is not clear whether this also applies to the time $t_\mathrm{max}$ at which the maximum of $\Delta w$ is reached,
but simulations suggest that this is the case (see fig.~\ref{Nmaxandtmax}).

\begin{figure}[htb]
\includegraphics[width=0.6\columnwidth]{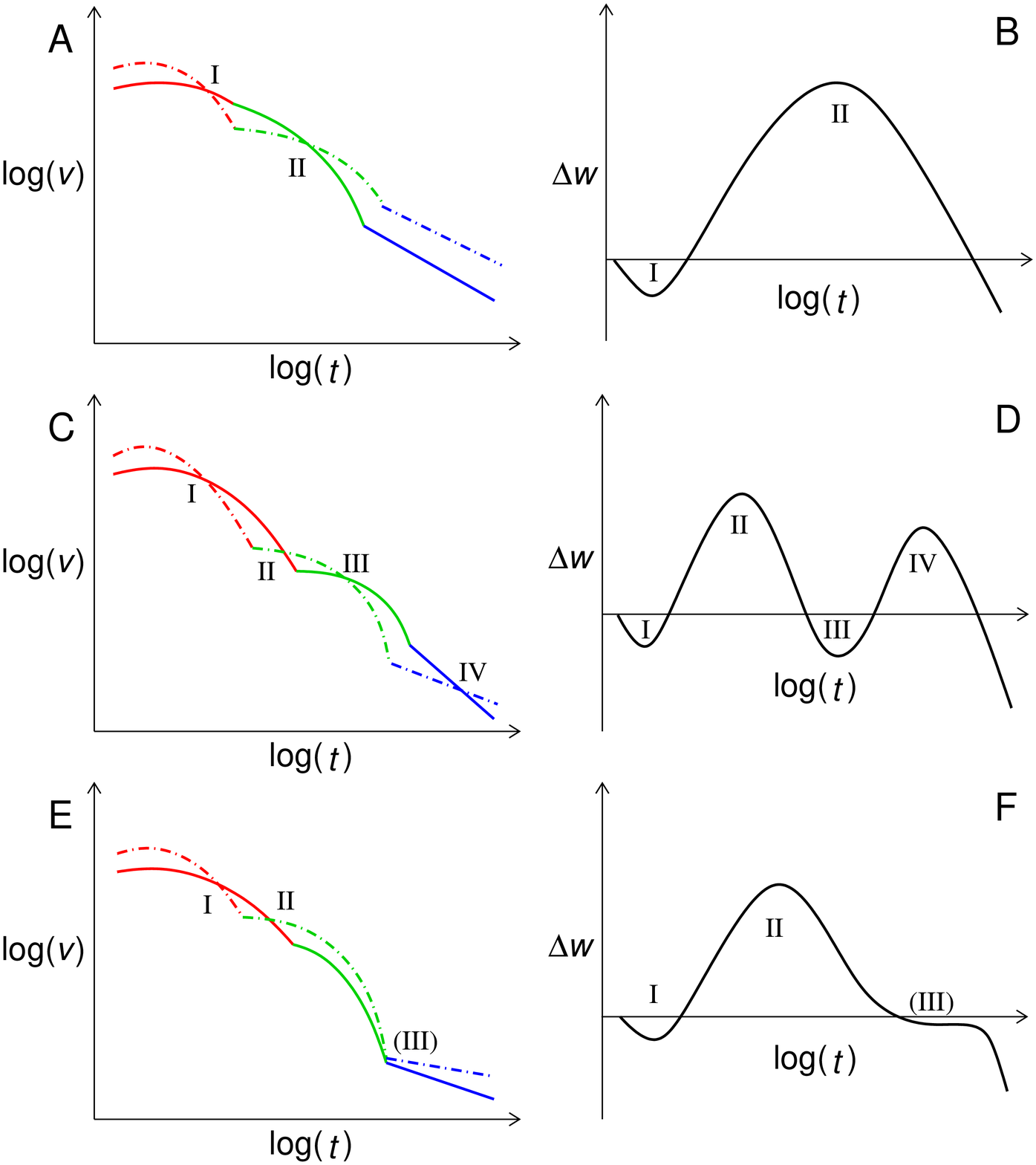}
\caption{\textbf{Summary of scenarios for a transient advantage of
    recombination.} The figure shows schematic plots of fitness velocities
    $v_\mathrm{r}$ (solid lines), $v_\mathrm{nr}$ (dashed lines) (A, C, E) and
    the corresponding behavior of $\Delta w$ (B, D, F).
Red, green and blue line segments correspond to the initial, ASD and final regime, respectively.
Latin numbers indicate intersections of the velocities and the
corresponding extrema in $\Delta w(t)$. The last number in panels E
and F is written in parenthesis to indicate that the two curves do not actually intersect but come very close to each other.
\label{veloc_schema}}
\end{figure}

In order to succinctly summarize the findings of the preceding
sections and indicate how they can be generalized, we rephrase them in
terms of the relative temporal ordering of the three different
regimes of the evolutionary dynamics. These regimes are:
 \begin{enumerate}
  \item The \textit{initial regime} where the population starts spreading from its monomorphic
 initial condition. Most of the fitter neighboring genotypes will be populated and this
 regime is marked by a large diversity.
 \item The regime of \textit{almost sequential dynamics (ASD)} where the population is quasi-monomorphic and moves by sequentially 
 fixing at increasingly fit, mostly neighboring, genotypes. It ends when a local maximum is reached.
 \item In the \textit{final regime}, the population is trapped at local maxima most of the time. Since
 escape is only possible by crossing a fitness valley, adaptation becomes very slow.
 \end{enumerate}
In practice it may be hard to clearly distinguish these regimes in an actual
fitness time series. For a systematic discussion it is convenient to introduce the fitness velocities defined as 
$v(t)=(\langle w(t+1)\rangle-\langle w(t-1)\rangle)/2$, again with
subscripts ``r'' and ``nr'' for recombining and non-recombining populations, respectively. The different
regimes correspond to different behaviors of $v$. In order to explain
the phenomenology we refer to a schematic picture of the velocities
and the corresponding behavior of $\Delta w$ in fig.~\ref{veloc_schema},
for plots showing real data see fig.~\ref{velo4intersec}.
Note that intersections of $v_\mathrm{r}$ and $v_\mathrm{nr}$ correspond to 
extrema in $\Delta w$. As a consequence, $\Delta w$ can have several maxima depending on the number
of intersections. In fig.~\ref{veloc_schema}
we display three cases, one with a single maximum followed by
a smooth decay, one case with two maxima, and finally again a case with a single maximum
but with a hump in the eventual decay.

%%%%%%%%%%%%%%%%%%%%%%%%%%%%%%%%%%%%%%%%%%%%%%%%%%%%%%%%%%%%%%%%%%%%%%%%%
%%%%%%%%%%%%%%%%%%%%%%%%%%%%%%%%%%%%%%%%%%%%%%%%%%%%%%%%%%%%%%%%%%%%%%%%%   

For both recombining and non-recombining populations, 
the velocities decay monotonically, apart from a short initial
period. This mainly reflects the simple fact that, in a static fitness
landscape, the availability of beneficial mutations decreases as the fitness level of the population
rises \cite{Park2008}.
This circumstance is aggravated by the structure of the RMF landscape, 
where after each step away from $\sigma^\ast$ the number of uphill
neighbors, which have a greater likelihood to be of higher fitness, 
is decreased by one \cite{neidhart2014}. In addition, as time elapses, in an increasing number of realizations the populations will end up 
trapped in local maxima, further decreasing the average velocity of the ensemble.

Non-recombining populations start increasing their fitness more rapidly at short times as they immediately 
concentrate on some particularly fit neighbors, while recombining populations keep a larger diversity that lowers 
their average fitness. Due to this concentration the non-recombining
populations start their ASD and therefore their velocity falls off rapidly, in accordance with Fisher's fundamental theorem. 
Because of their higher diversity the velocity of the recombining
populations falls off more slowly, allowing them to overtake the non-recombining ones at least in velocity.

When all the non-recombining populations either became trapped or entered the ASD, 
the decay of their average velocity slows down, while for the recombining populations the decay is still 
rapid. The subsequent behavior depends on whether the recombining populations on average enter their ASD regime before 
or after their velocities have fallen below the level of the
non-recombining populations. The first case is illustrated in
fig.~\ref{veloc_schema}A. 
Eventually, as all populations get trapped on local optima, the velocity of 
recombining populations must fall below that of non-recombining ones, giving rise to a second intersection 
and thus to a single maximum in $\Delta w$ (cf.\ fig.~\ref{veloc_schema}B).

On the other hand, if recombining populations enter the ASD regime
after the point where $v_\mathrm{r}(t) = v_\mathrm{nr}(t)$, two
scenarios are possible. 
Either, due to the deceleration when the recombining populations enter
the ASD regime, a third intersection emerges which gives rise to a
second maximum in $\Delta w$ (fig.~\ref{veloc_schema}C,D). Or there
is no third intersection, but due 
to the slower velocity decay, the velocities of recombining and
non-recombining populations approach each other, causing a hump in the 
decay of $\Delta w$ after the maximum (fig.~\ref{veloc_schema}E,F).  
Simulation data illustrating all three scenarios are displayed in fig.~\ref{velo4intersec}.

\subsection*{Fitness seascapes\label{sec:seascape}}
If recombination is only of transient advantage one must ask why it is
nevertheless so ubiquitous. One explanation may be that recombination would preferably show up in the case of weak selection, i.e.\ nearly neutral fitness landscapes. In such cases, recombination can break up linkage, impeding trappings, so that the populations can be expected to behave similar to populations on smooth landscapes \cite{neher2009,neher2013}. 

If recombination is also ubiquitous in the presence of strong selection, there must exist mechanisms that considerably prolong the period of advantage.
For example, as we have shown earlier, the larger the number of loci,
the more advantageous recombination is and the longer the advantage
lasts. Another mechanism that might prolong the advantageous period is
disruptive selection \cite{Rueffler2006}, 
which can lead to stable coexistence of distant genotypes. This would
make genetically homogeneous populations less 
likely and would thus reduce the probability of trappings at local optima. 

Moreover, a prolonged beneficial effect of recombination can result from the genome being organized in such ways that epistasis, 
which may give rise to local fitness maxima, only occurs within fixed
blocks of the genome which are preserved under recombination \cite{watson2011}. 
Strong trapping due to recombination can only occur at local maxima that must be escaped by combining mutations on loci located on different blocks. 
As in the setup proposed in \cite{watson2011} no such local maxima exist, the advantage of recombination should prevail forever. 
However, more realistically, there would be at least some epistasis
also between the blocks. For large genomes the existence of local
optima would then still be likely, although the times for which the
population can adapt freely before getting caught in such a state may be extremely long.

Here we examine a fourth possibility to prolong the benefit of recombination. 
Suppose a population lives in an environment that changes rapidly in time. 
Then the probability for the population to survive does not depend so much on its ability to reach the fittest possible state in the long run, but rather to rapidly escape from the poorly adapted state it finds itself in after the external conditions have abruptly changed. 
Therefore, we expect recombination to be beneficial in the long run if the time scale at which the fitness landscape changes is shorter than the time scale at which recombination stops being beneficial on a static landscape.
Such fluctuating fitness landscapes are sometimes called fitness seascapes \cite{mustonen2009}. 
Note that a resetting of a fitness landscape does not necessarily need to be the result of fluctuating environments, but
could also be the result of a population traveling through a spatially inhomogeneous world.  

In order to test our hypothesis that the beneficial effect of recombination may be sustained indefinitely in fluctuating environments we constructed seascapes as follows. 
At each time step of the population dynamics, the landscape is reset with probability $p_\mathrm{r}$, i.e.,
the random variables $X_\sigma$ defining the random fitness components in eq.~(\ref{eqn:def_landscapes}) 
are redrawn and the reference sequence $\sigma^\ast$ is newly chosen. 
The new reference sequence is either selected from one of the neighbors of the preceding one
or picked completely at random. We will refer to the former selection mode as a \textit{soft reset} and to
the latter as a \textit{hard reset}. 
Note that models of single-peaked
fitness landscapes with a moving optimum have been considered
previously \cite{maynardsmith1988,charlesworth1993,kondrashov1996,nilsson2002,Wilke2006} (see also \cite{Buerger2000,Otto2009} and references therein). 

%%%%%%%%%%%%%%%%%%%%%%%%%%%%%%%%%%%%%%%%%%%%%%%%%%%%%%%%%%%%%%%%%%%%%%%%%%%%%%%%%%%%%%%%%%%%%%%%%555
\begin{figure}[htb]
\includegraphics[width=0.5\columnwidth]{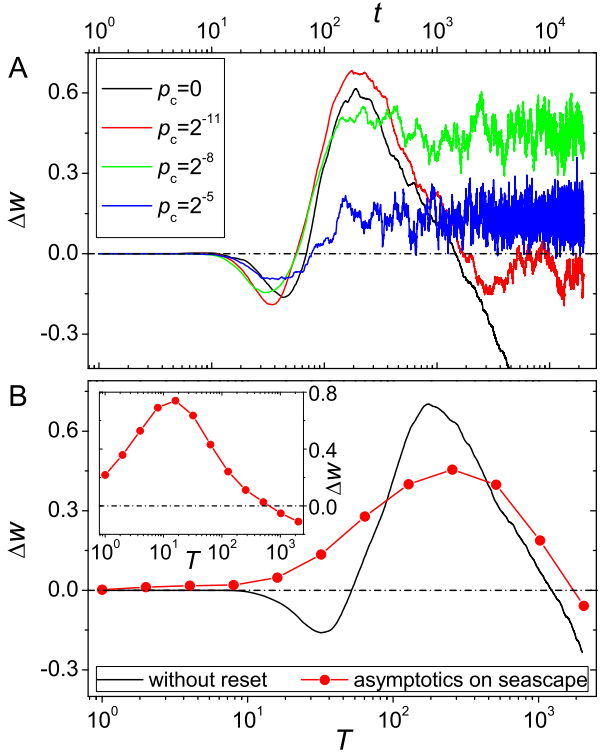}
 \caption{\textbf{Advantage of recombination in fitness seascapes.} (A) Time series of the fitness advantage $\Delta w$ of a seascape
with hard reset. (B) Asymptotic fitness advantage $\Delta w$ in dependence 
on $T=1/p_\mathrm{r}$ in the stationary state of the seascape
with hard reset. The data are extracted from the time series and correspond to the value of $\Delta w$ 
averaged over generations 5000 to 20000.
For comparison we also plotted the time series
for the same set of parameters on the corresponding (static) landscape
and a random initial genotype.
The parameters are $c=0.75$, $N=2000$ and $\mu=0.0025$. In the inset we
show the stationary advantage on a seascape with soft reset.\label{fig:timeseries_seascape}}
\end{figure}
%%%%%%%%%%%%%%%%%%%%%%%%%%%%%%%%%%%%%%%%%%%%%%%%%%%%%%%%%%%%%%%%%%%%%%%%%%%%%%%%%%%%%%%%%%%%%%%%%5

Since local maxima only exist temporarily in this setting, the population will
not be trapped forever at any specific genotype.
Instead, the global maximum
will be continuously pursued.
In case of a soft reset, the direction of the global fitness gradient remains essentially unchanged 
after resetting the seascape whereas the adaptation process starts anew from a random position in case of a hard reset.
In both cases the mean fitness advantage of recombination
becomes stationary after some
time (fig.~\ref{fig:timeseries_seascape}A). The mean value of this stationary advantage depends on how well the
population is able to follow the global optimum. Based on our results for static landscapes, 
it is clear that the recombining population has the largest
advantage if the time $T_\mathrm{r}=1/p_\mathrm{r}$ between two resets
of the seascape is of the same order as the time $t_\mathrm{max}$ that a population 
on a static landscape can evolve freely before being trapped at a local maximum.
Moreover, recombination becomes disadvantageous if $T_\mathrm{r} > t_0$,
i.e., if the time between two resets is longer than the time a non-recombining population needs to overtake a
recombining population.

As a consequence, the asymptotic fitness advantage $\Delta w$
with respect to $T_\mathrm{r}$ on a seascape behaves similar to the time-dependent advantage 
$\Delta w(t)$ on a static landscape (fig.~\ref{fig:timeseries_seascape}B). Using a physical analogy, we can say
that the stationary advantage of recombination is maximized when the intrinsic time scale of adaptation is in resonance  
with respect to the environmental alteration time.
The dip at short times in the time series $\Delta w(t)$ is not visible in the stationary advantage $\Delta w(T_\mathrm{r})$.
This is because, after the resetting of the landscape, the population is not monomorphic, which means
that the short time regime associated with the buildup of diversity discussed previously in the context of static
landscapes is absent on the seascape.
Note also that the curves corresponding to the time-dependent advantage on static landscapes and the stationary advantage on the 
seascape coincide only if suitable initial conditions
are used in the former case. For the comparison with the hard reset in fig.~\ref{fig:timeseries_seascape}B,
the initial population in the static landscape simulations was therefore placed at a randomly
chosen sequence rather than at $\sigma^\ast$.

\section*{Discussion}

In this article we have numerically examined the benefit of
recombination in a class of tunably rugged fitness landscapes. A few
earlier papers have been devoted to the systematic study of the effect of sign epistasis on the advantage of sex on high
dimensional fitness landscapes \cite{kondrashov2001,watson2005,devisser2009,bonhoeffer2009,moradigaravand2012}, but they
either considered very specific landscape realizations \cite{kondrashov2001,watson2005,devisser2009} or did not take
into account the temporal development of the evolutionary advantage due to recombination
\cite{bonhoeffer2009,moradigaravand2012}. Focusing on this temporal development, we find that the advantage of
recombination is a strictly transient effect, which is particularly interesting as this observation coincides nicely
with recent experiments carried out with the facultatively sexual rotifer \textit{Brachionus calyciflorus}
\cite{becks2012}. We note that an increased appreciation of transient, as
compared to asymptotic mechanisms has also been promoted in the context of ecological
theory \cite{Hastings2004}.                           

Our analysis suggests that 
several of the well known effects
associated with recombination play an important part in the evolutionary dynamics,
but their importance differs at different timescales.
At very short timescales the Weismann
effect is dominant, forcing the population to distribute more broadly
over the genotype space. 
This leads to an initial fitness disadvantage of recombination, but note that this is to some
extent an artifact of the monomorphic initial condition. Such a short time disadvantage has been reported in
experiments with \textit{Saccharomyces cerevisiae} \cite{greig1998}. Although there the initial disadvantage of
recombination was attributed to the costs of sexual reproduction which are absent in our numerical setup, it is 
plausible that the disadvantage is at least partly caused by the mechanism we described here.

At intermediate timescales, the Fisher-Muller effect makes use of the
increased population diversity for producing distant high-fitness mutants, which leads to an advantage of recombination.
This regime continues until the populations concentrate at local
fitness maxima. Using known results from studies of 2-locus-models, we argue that, especially for large $N$, recombination prevents the escape from local optima and populations
get trapped on such states for very long times. For non-recombining
populations the trapping is much weaker, leading to shorter escape
times. Hence they can proceed gaining fitness while recombining
populations remain stuck, losing all the previously built up advantage
in the long run. 
With an increasing number of optima, such trappings will of course
occur earlier. 
Roughly speaking, the advantage of recombination
lasts longer the less rugged the underlying landscape is.

An important question that we have discussed is how relevant such
trapping events are in very large landscapes corresponding to
thousands of genes or millions of nucleotides. For a state to be a
local maximum, all its neighbors must be of lower fitness, which is
the more improbable the larger the number of loci $L$. This question is
not easy to address in general terms, as the answer will depend on the
landscape model \cite{franke2011}. Referring to known results for
adaptation in the SSWM regime 
\cite{gillespie1983,flyvbjerg1992,orr2003,krugneid,jain2011}, we have
argued that trapping will remain relevant provided the distance
travelled by an adaptive walk to a local optimum increases
sufficiently slowly (logarithmically) with the number of loci, as
appears to be the case for the RMF model   
provided the slope parameter $c$ is not much larger than the
ruggedness parameter $\lambda$ \cite{neidhart2014}.  
Thus the scenario proposed here remains valid for large $L$ if the
fitness landscape is sufficiently rugged. 

The outstanding importance of the strong trapping of recombining
populations has previously been noted in \cite{moradigaravand2012},
where also the advantage of recombination on high dimensional fitness
landscapes with epistasis was studied. Unlike in our study, however,
the advantage of recombination was evaluated after a fixed time. The
same is true for \cite{bonhoeffer2009}, where different quantities
characterizing the landscape were tested with respect to their ability
to predict whether recombination will be advantageous or not. Our
results show that the question of whether recombination is
advantageous is well defined only if the corresponding temporal scale is specified.  
Unless the time scale is clearly determined by the context, e.g., for
a particular experimental setup or an ecological scenario, it may be
more appropriate to assess the advantage of recombination through
quantities like $t_0$, $t_\mathrm{max}$ or $\Delta w
(t_\mathrm{max})$ defined above. 

A transitional benefit of recombination has been observed in various
previous studies, both on the level of the fitness trajectories of
recombining vs. non-recombining populations \cite{kim2005} and with
regard to the temporal change of the frequency of a modifier allele
governing the rate of recombination (see e.g.\ \cite{Feldman1997,Otto2009}). 
However, these studies did not consider fitness landscapes with local
maxima where populations could get trapped, and the transient nature
of the recombinational advantage arose because
the global maximum was found by the recombining population. 
This difference is important when the number of loci
becomes large, because then finding the global optimum is illusive and
thus irrelevant for the dynamics, while getting trapped by local maxima is still very common.

We have shown that on time dependent landscapes the advantage
of recombination can subsist for indefinite times as long as the time
scale on which the landscape changes is shorter than the time scale at
which non-recombining populations can overtake trapped recombining
ones. 
This finding is in concordance with van Valen's Red Queen Hypothesis
\cite{valen1973}, 
and it specifies the time scale of environmental change that confers
the maximal long-time benefit. 
An indefinitely prolonged advantage of recombination on time dependent
landscapes with a single moving optimum has been reported previously
(see e.g. \cite{Buerger2000} and references therein). Again, our model emphasizing the importance
of trapping at local optima provides a rather different perspective on
this effect,  
because the time scale on which local optima appear and vanish 
can be significantly different from the dynamics of the global peak.

A number of interesting questions could not be addressed in this study.
For example, this article deals only with the dynamics in limit where selection is strong compared to recombination.
In \cite{neher2009,neher2013} it has been shown that in the case of weak selection,
a different dynamics of recombining populations is to be expected, where recombination can break up linkage,
such that trappings do not occur anymore and the populations are well described by their allele frequencies.
Although one might expect that in this regime the populations behave similarly to those on smooth landscapes,
this point needs to be investigated. Further, it would be important to determine in experiments which of the two
regimes is the relevant one for naturally recombining populations.   

Another open questions concerns the behavior of diploid populations.
Considering the case of haploids, the existence of appropriate local
maxima yields multiple equilibria in the $N \rightarrow\infty$ limit,
where the recombining populations are strongly centered around one of the suboptimal peaks.
Strictly speaking, these are not stationary states for finite $N$,
but the times that a recombining population needs to escape from such states can still be extremely long.
For diploid populations in the $N \rightarrow\infty$ limit, on the other hand,
multiple equilibria can exist even in the absence of local maxima in the genotypic fitness landscape\cite{buerger1989}.
It would therefore be interesting to investigate, whether the existence of such multiple
equilibria that emerge in the absence of sign epistasis will lead to long trapping times for
finite diploid populations.  

Finally, it might also be important to investigate other, perhaps more realistic, recombination schemes such as single- or multi-point
crossovers. Another interesting choice would be block recombination \cite{watson2011}, which, for suitable epistatic interactions within or between blocks, has already been
shown to be able to considerably prolong the time for which recombination is beneficial.

\section*{Models and methods \label{sec:model}}
\subsection*{Fitness landscape}

We consider haploid populations with a binary alphabet for the
representation of the genome, i.e.\ the genome is modeled as a binary
sequence $\sigma = (\sigma_1, \sigma_2, ..., \sigma_L)$ of a fixed length $L$. This sequence can for example be
interpreted as $L$ genes or nucleotides that are either identical to a
wild type allele or a mutated variant. 
The set of genotypes is then given by the hypercube $\mathbb{H}_2^L=\{0,1\}^L$.
Together with the Hamming distance 
\begin{equation}
\label{Hamming}
d(\sigma, \tau)=L - \sum_i \delta_{\sigma_i \tau_i}
\end{equation}
between two genotypes $\sigma, \tau \in \mathbb{H}_2^L$
this structure becomes an $L$-dimensional metric space, the Hamming space.
In eq.~(\ref{Hamming}) the Kronecker symbol equals $\delta_{x,y} = 1$ if $x = y$ and 0 else. 

To each genotype $\sigma$ we assign a real number $w(\sigma)$ that 
represents the fitness associated with $\sigma$. 
In this paper we mostly consider dynamics on landscapes created
according to the Rough Mount Fuji (RMF) model
\cite{aita2000,aita2000a,franke2011,neidhart2014}, where the fitness values are assigned as
\begin{align}
 w(\sigma) = X_\sigma + c\,d(\sigma, \sigma^\ast).
\label{eqn:def_landscapes}
\end{align}
Here the $X_\sigma$ are  random
numbers drawn \textit{independently} from an
exponential distribution with mean $\lambda$, $c$ is the mean slope of the landscape, and $\sigma^\ast$ is a reference sequence.
On average, the fitness values become larger with increasing distance to $\sigma^\ast$. 
For $\lambda=0$ the landscape defined by (\ref{eqn:def_landscapes}) is purely additive,
and for $c=0$ it reduces to a model where fitness values are assigned randomly and independently to genotypes \cite{kaufflev}. 
Because of its similarity to a mutation-selection model introduced in \cite{kingman2} the latter case is often 
referred to as the House of Cards (HoC) model \cite{deVisser2014,franke2011,Szendro12}.  
To test the
robustness of our conclusions we present results obtained using
Kauffman's NK-model \cite{kauffman1989} in fig.~\ref{lk_kdep} (see
\cite{franke2011,Franke2012,Schmiegelt2014} for further
information about this model). 

\subsection*{Finite population dynamics}

We use a slightly modified version of the Wright-Fisher model with constant
population size. 
Let $N$ ($N_\sigma$) denote the number of individuals in the population (carrying genotype $\sigma$) and $f_\sigma = N_\sigma / N$
the fraction of individuals with genotype $\sigma$.
A single time step of the population dynamics consists of the following substeps:

\begin{description}
 \item[Mutation:] On each genotype $\sigma$, a fraction $\mu$ of individuals mutates to the neighboring genotypes:
$$ f_\sigma \,\to\, (1-\mu) f_\sigma + \frac{\mu}{L} \sum_{\underset{d(\sigma,\tau)=1}{\tau}} f_\tau \,.$$

 \item[Selection:] The fraction of individuals on each genotype is decreased or increased, depending
on the fitness value compared to the average fitness:
$$ f_\sigma \,\to\, \frac{w(\sigma)}{\overline w} f_\sigma $$
where $\overline w = \sum_\sigma f_\sigma w(\sigma) $. 

 \item[Random sampling:] On each genotype $\sigma$, the number of individuals is replaced by a random variable drawn from a Poisson distribution
with mean value $N_\sigma=f_\sigma N$. After this process, the population size is 
actually $N + \mathcal{O}(\sqrt{N})$, such that we have to normalize the population
in order to keep it constant. Note that usually a multinomial distribution on
all genotypes at once is used instead of independent Poisson distributions. However,
we chose this method because it makes the simulation much faster \cite{zanini2012}.

 \item[Recombination:] A fraction $r$ of individuals is replaced by other individuals
whose genotypes are recombinants of two randomly chosen parents. Recombination
is performed with a uniform crossover scheme, i.e., each locus is either
taken from the first or the second parent with equal probability. If $N r \le 1$,
the number of individuals which is replaced is drawn from a Poisson distribution
with mean value $N r$.
\end{description}

\subsection*{Infinite population dynamics}\label{sec:infiniN}

The infinite population size limit results in a deterministic dynamics
on the set of genotype frequencies $f_\sigma(t)$ \cite{Buerger2000}. For convenience we employ here an algebraic
formulation, which is explained in the following.

A single mutation can only transfer a sequence into one of its
neighbors. The \textit{adjacency matrix} encodes the
information about the neighborhood structure of the hypercube, and is defined by 
\begin{align}
    \mathcal A_{\sigma, \sigma'} = \begin{cases} 1, \, \text{ if } d(\sigma, \sigma') = 1,\\
                               0, \, \text{else}. \end{cases} \nonumber
\end{align}                      
Considering an evolutionary dynamics with mutation probability $\mu$,
at every time step a fraction $(1-\mu)$ of the population with a given
genotype does not mutate, while
a fraction $\frac \mu L$ mutates to each of its $L$ neighbors. Thus, the single step mutation matrix is defined
by 
\begin{align*}
    \mathcal M ^{(1)} = (1-\mu) \mathbb I + \frac \mu L \mathcal A \nonumber
\end{align*}
where $\mathbb I$ denotes the identity operator. 
As $N$ gets larger, double mutations that appear with a rate of order
$\mu^2$ become more frequent and must be taken into account. The mutation matrix including two step mutation events is 
\begin{align*}
    \mathcal M^{(2)} =  (1-\mu) \mathbb I + \mathcal M^{(1)}
    \frac{\mu}{L} \mathcal A =   (1-\mu) \mathbb I  + 
\left(  (1-\mu) \mathbb I +\frac \mu L \mathcal A\right)\frac \mu L
\mathcal A. \nonumber
\end{align*}
In the same manner, this can be generalized to include mutations up to $n$--th order:
\begin{align}
    \mathcal M^{(n)} &= (1-\mu) \mathbb I + \mathcal M^{(n-1)} \frac
    \mu L \mathcal A \nonumber \\ &=   (1-\mu) \mathbb I  + \left( (1-\mu) \mathbb I
      +\left(  (1-\mu) \mathbb I
    +\dots \frac \mu L \mathcal
    A\right)\dots\right)\frac \mu L \mathcal A \nonumber \\
    &=\sum_{j=0}^n(1-\mu)(\frac \mu L \mathcal A)^j
    \stackrel{n\to \infty}{\longrightarrow} (1-\mu)\left(\mathbb I
      -\frac \mu L \mathcal A\right)^{-1}=\mathcal M^{(\infty)}, \nonumber
\end{align}
where the limit is performed using the geometric series of matrices, which exists because $\lim_{k\to \infty}(\frac \mu L\mathcal A)^k =0$. 
Note that for infinite populations mutations of no order can be neglected and, thus, $\mathcal M^{(\infty)}$ is the
appropriate mutation matrix in this limit.

Recombination still only happens once per generation, thus it is
realized by the corresponding transition matrix $\mathcal T$ defined in \cite{stadler1998}.
The selection matrix is 
\begin{align}
\mathcal S_{\sigma,\sigma'} = w(\sigma)\delta_{\sigma, \sigma'}. \nonumber
\end{align}
Then the genotype frequencies $f \in [0,1]^{2^L}$ evolve
in time according to
\begin{align}
f_\sigma(t+1)=\frac{1}{\sum_{\sigma'\in\mathbb H} \left(\mathcal{T S
  M^{(\infty)}}f\right)_{\sigma'}(t)} \left( \mathcal{TSM^{(\infty)}}f\right)_\sigma(t).\nonumber
\end{align}
This means in particular, that already after the first time step every genotype is created by mutation. Starting with a monomorphic
population, $\mathcal M^{(\infty)}$ leads to a distribution of the population that decays exponentially with the Hamming distance to the initially populated state.
This can cause numerical problems if the numbers representing the population become too small.
In order to be sure that the population is not neglected on any state of the landscape at any time due to limits of computational precision,
the simulations were performed with a precision of more than 30 significant digits
and we restrict ourselves to small numbers of loci.

\section*{Acknowledgements}
We are grateful to Richard Neher for comments on an earlier version of
the manuscript. This work was funded by Deutsche
Forschungsgemeinschaft through the Bonn Cologne Graduate School for Physics and Astronomy and through grants SFB 680, SFB TR12 and SPP 1590.

%%%%%%%%%%%%%%%%%%%%%%%%%%%%%%%%%%%%%%%%%%%%%%%%%%%%%%%%%%%%%%%%%%%%%%%%%
%For supplementary material
\newpage

% \newpage

\section*{Supporting Figures}
 
\setcounter{figure}{0}
\renewcommand{\thefigure}{S\arabic{figure}}

\pagestyle{empty}

\vspace*{1.cm}

% \clearpage
\begin{figure}[h!]
\includegraphics[width=0.8\columnwidth]{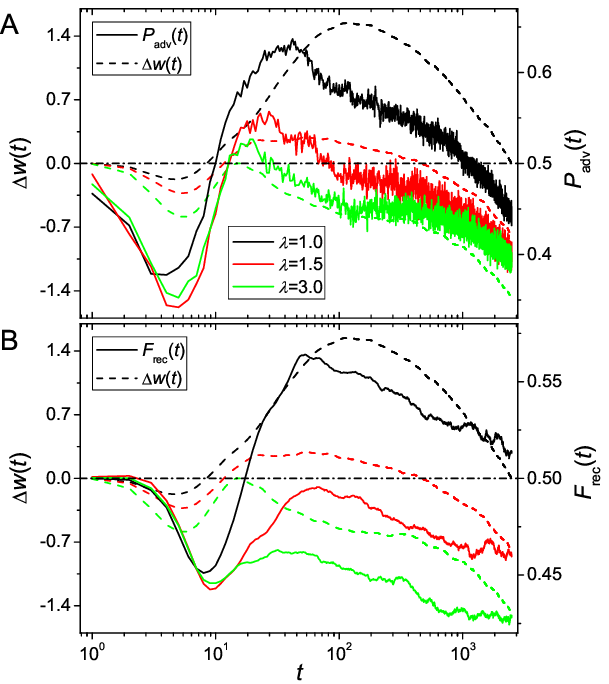}
\caption{Comparison of $\Delta w(t)$, $P_\text{adv}(t)$ and the frequency of
recombining individuals for the case with a modifier allele $F_\text{rec}$ as quantifiers for the
(dis-)advantage
of recombination. Parameters are the same as in fig.~\ref{deponlambda}.
All quantities are correlated and show qualitatively
the same behavior. Nevertheless, there are regions on the time axis where
$\Delta w$ shows a recombination advantage while $P_\text{adv}(t)$ indicates
a disadvantage and vice versa. This implies that the distribution of
$\Delta w$ is not centered around the mean value, or more precisely, the
mean value is not equal to the median.
\label{padv}
}
\end{figure}
%%%%%%%%%%%%%%%%%%%%%%%%%%%%%%%%%%%%%%%%%%%%%%%%%%%%%%%%%%%%%%%%%%%%%%%%

\begin{figure}[h!]
\includegraphics[width=0.8\columnwidth]{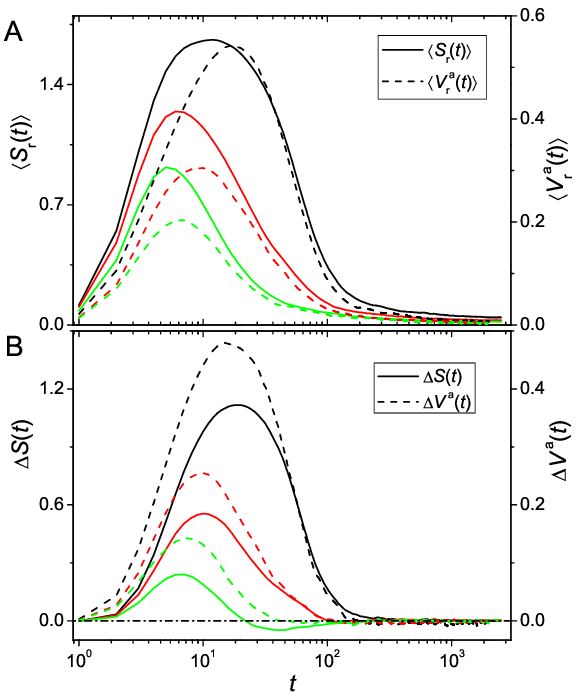}
\caption{
  \label{Sandvarn}
  Comparison of the entropy $S_r$ and the additive genetic variance $V^a_r$ (A)
  as well as $\Delta S$ and $\Delta V^a$ (B). 
Parameters are the same as in fig.~\ref{deponlambda}.
All quantities are correlated and show qualitatively
the same behavior. }
\end{figure}

\begin{figure}[h!]
\includegraphics[width=0.8\columnwidth]{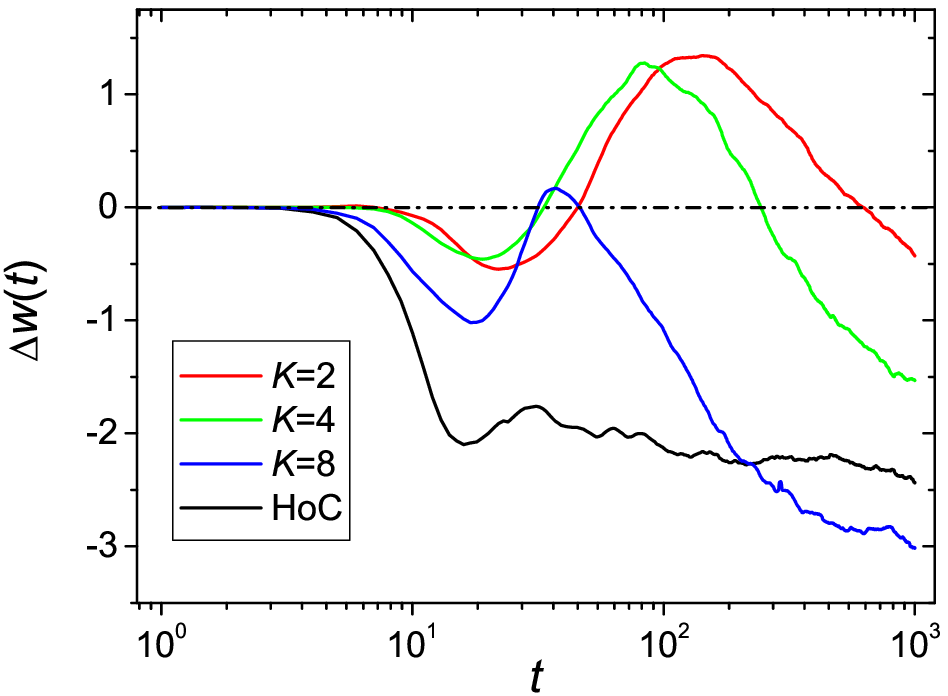}
\caption{
Transitory advantage of recombination on Kauffman's NK-landscape with $L=16$ binary loci. Each locus
interacts with $K$ adjacent neighbors, and fitness values are drawn from a lognormal distribution. 
Population parameters are $N=1000$ and $N\mu=2$. For increasing
ruggedness, i.e., for increasing values of $K$, the advantage becomes
less pronounced and vanishes. Data marked `HoC' correspond to the
maximally rugged case $K=L-1=15$, where the NK-model reduces to the
House of Cards model with uncorrelated random fitness values. 
\label{lk_kdep}
}
\end{figure}

 %%%%%%%%%%%%%%%%%%%%%%%%%%%%%%%%%%%%%%%%%%%%%%%%%%%%%%%%%%%%%%%%%%%%%%%%%
\begin{figure}[h!]
\includegraphics[width=0.8\columnwidth]{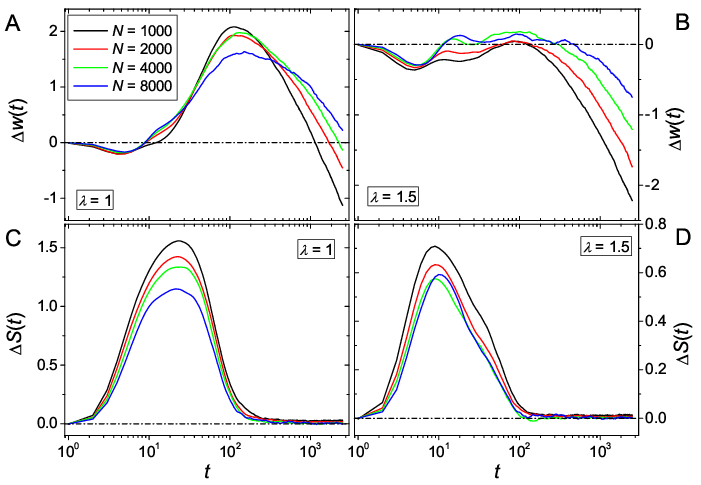}
\caption{Dependence of the recombinational advantage on
    population size. Same as figure~\ref{deponNmu} but with constant $N\mu = 8$ and
varying population size $N$.\label{deponN}}
\end{figure}
%%%%%%%%%%%%%%%%%%%%%%%%%%%%%%%%%%%%%%%%%%%%%%%%%%%%%%%%%%%%%%%%%%%%%%%%

%%%%%%%%%%%%%%%%%%%%%%%%%%%%%%%%%%%%%%%%%%%%%%%%%%%%%%%%%%%%%%%%%%%%%%%%%
\begin{figure}[h!]
\includegraphics[width=0.8\columnwidth]{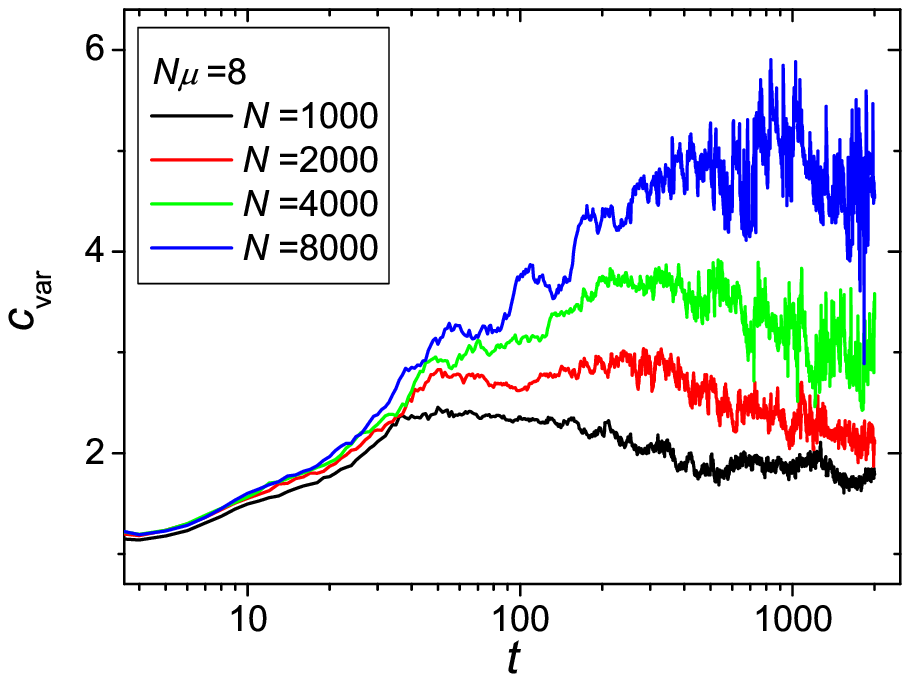}
\caption{The coefficient of variation $c_\mathrm{var}$ of the number of mutants that are not located at the most populated genotype when the latter is a local optimum vs.\ $t$ for systems with $r=1$, $c=1$, and $\lambda=1$. Note that, for constant $N\mu$, $c_\mathrm{var}$ increases with $N$.\label{cvar}}
\end{figure}
%%%%%%%%%%%%%%%%%%%%%%%%%%%%%%%%%%%%%%%%%%%%%%%%%%%%%%%%%%%%%%%%%%%%%%%%

%%%%%%%%%%%%%%%%%%%%%%%%%%%%%%%%%%%%%%%%%%%%%%%%%%%%%%%%%%%%%%%%%%%%%%%%
\begin{figure}[h!]
\includegraphics[width=0.8\columnwidth]{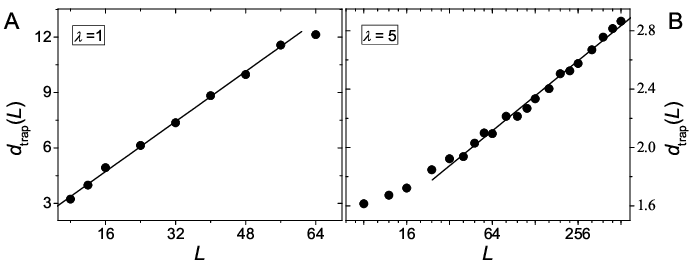}
\caption{Trapping distance increases with landscape
    dimensionality. The figure shows the distance $d_\mathrm{trap}$ at which populations are trapped for the first time vs. number of loci $L$ for $N=2000$, $N\mu=4$, $c=1$, and (A) $\lambda=1$, (B) $\lambda=5$. Straight lines are guides to the eye and show that the $L$-dependencies are well compatible with a (A) linear and (B) logarithmic relation, respectively.
\label{fig:Ldeptrapn}}
\end{figure}
%%%%%%%%%%%%%%%%%%%%%%%%%%%%%%%%%%%%%%%%%%%%%%%%%%%%%%%%%%%%%%%%%%%%%%%%
% \clearpage

% \newpage
%%%%%%%%%%%%%%%%%%%%%%%%%%%%%%%%%%%%%%%%%%%%%%%%%%%%%%%%%%%%%%%%%%%%%%%%%
\begin{figure}[h!]
\includegraphics[width=0.8\columnwidth]{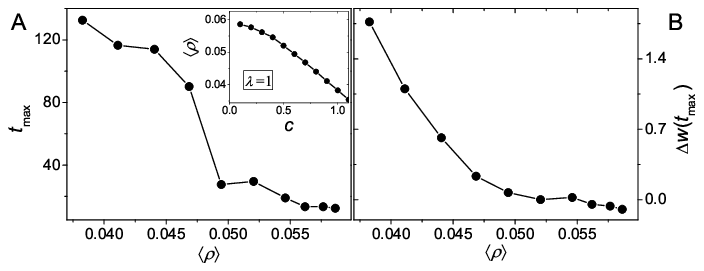}
\caption{Recombinational advantage declines with the density
    of local maxima. 
To clarify whether the time of maximal recombination advantage, $t_\mathrm{max}$, increases with decreasing density
of local maxima, $\rho$, we measured $\Delta w(t_\mathrm{max})(\langle\rho\rangle)$ and
$t_\mathrm{max}(\langle\rho\rangle)$ in simulations, where $\langle\rho\rangle$ is the mean
density of maxima averaged over realizations as well as over genotype
space (recall that the density of maxima is inhomogeneous in the RMF
model). Panel (A) shows $t_\mathrm{max}$ and panel (B) the corresponding maximal advantage $\Delta w(t_\mathrm{max})$ vs.
$\langle\rho\rangle$. Both quantities decline monotonically with increasing $\langle\rho\rangle$. The density was controlled by changing the slope $c$ for fixed $\lambda=1$ (see inset of panel (A)). 
 Data correspond to $N=1000$, $N\mu=4$ and $\lambda=1$. \label{Nmaxandtmax}}
\end{figure}
%%%%%%%%%%%%%%%%%%%%%%%%%%%%%%%%%%%%%%%%%%%%%%%%%%%%%%%%%%%%%%%%%%%%%%%%

\begin{figure}[h!]
\includegraphics[width=0.8\columnwidth]{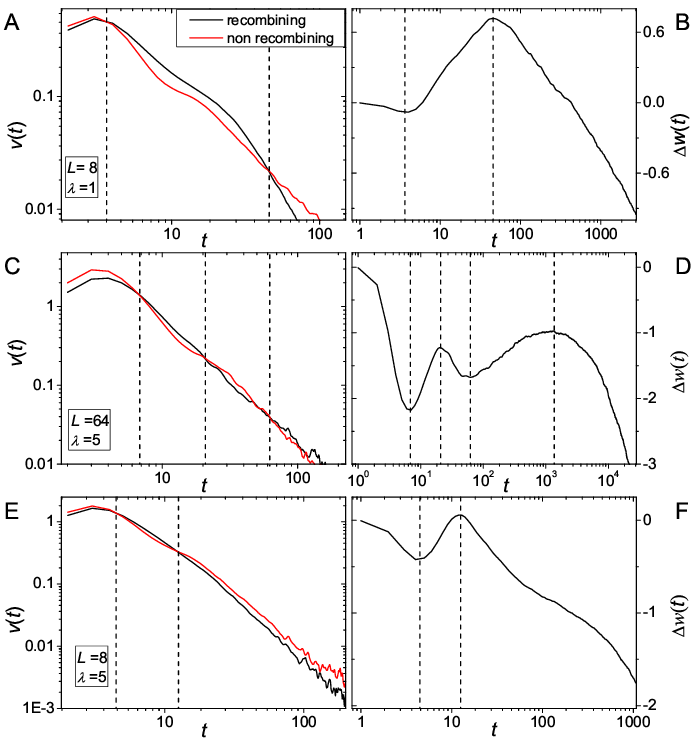}
\caption{(A), (C), and (E): Fitness velocities vs.\ $t$ for systems with parameters $N=2000$, $N\mu=4$, and $c=1$. Intersections are marked by dashed lines. (B), (D), and (F): Corresponding curve for $\Delta w$ vs.\ $t$. Compare with curves in the schematic pictures in fig.~\ref{veloc_schema}.\label{velo4intersec}}
\end{figure}
%%%%%%%%%%%%%%%%%%%%%%%%%%%%%%%%%%%%%%%%%%%%%%%%%%%%%%%%%%%%%%%%%%%%%%%%  

\vspace*{5.cm}

\end{document}